\documentclass[]{aa} 
\usepackage{graphicx}
\usepackage{natbib} 
\begin{document}
\title{The Wendelstein Calar Alto Pixellensing Project (WeCAPP):\\ The
M31 Variable Star Catalogue \thanks{Based on observations obtained at
the Wendelstein Ob\-ser\-va\-tory of the Universit\"atssternwarte
M\"unchen.}}

\author{J\"urgen Fliri\inst{1}\thanks{Visiting astronomer at the
German-Spanish Astronomical Center, Calar Alto, operated by the
Max-Planck-Institut f\"ur Astronomie, Heidelberg, jointly with the
Spanish National Commission for Astronomy.}, Arno
Riffeser\inst{1,2}$^{\star\star}$, Stella Seitz \inst{1}, \and~Ralf
Bender\inst{1,2}}
   
\offprints{fliri@usm.uni-muenchen.de}
   
\institute{Universit\"atssternwarte M\"unchen, Scheinerstrasse 1,
81679 M\"unchen\\ \and Max-Planck-Institut f\"ur Extraterrestrische
Physik, Giessenbachstrasse, 85748 Garching}
   
\date{retrieved, accepted: to be inserted later}

\abstract{In this paper we present the WeCAPP variable star catalogue
towards the bulge of M31. The observations of the WeCAPP microlensing
survey (optical $R$ and $I$ bands) during three years (2000-2003)
result in a database with unprecedented time coverage for an
extragalactic variable star study.  We detect 23781 variable sources
in a $16.1\arcmin \times 16.6\arcmin$ field centered on the nucleus of
M31.  The catalogue of variable stars contains the positions, the
periods and the variations in the $R$ and $I$ bands. We classify the
variables according to their position in the $R$-band period-amplitude
plane. Three groups can be distinguished; while the first two groups
can be mainly associated with Cepheid-like variables (population I
Cepheids in group I, type II Cepheids and RV Tauri stars in group II),
the third one consists of Long Period Variables (LPVs).  We detect 37
RV Tauri stars and 11 RV Tauri candidates which is one of the largest
collections of this class of stars to date.  The classification scheme
is supported by Fourier decomposition of the light curves. Our data
shows a correlation of the low-order Fourier coefficients $\Phi_{21}$
with $\Phi_{31}$ (as defined by \citet{simon81}) for classical
Cepheids, as well for type II Cepheids and RV Tauri stars.
Correlating our sample of variable stars with the X-ray based
catalogues of \citet{kaaret02} and \citet{kong02} results in 23 and 31
coincidences, 8 of which are M31 globular clusters. The number density
of detected variables is clearly not symmetric, which has to be
included in the calculations of the expected microlensing event rate
towards M31.  This asymmetry is due to the enhanced extinction in the
spiral arms superimposed on the bulge of M31 which reduces the number
of sources to about 60\%, if compared to areas of equivalent bulge
brightness (without enhanced extinction being present).
     
}
  
\titlerunning{WeCAPP variable stars} \authorrunning{Fliri et al.}
   \maketitle
%
%
\section{Introduction}
In the last decade the ongoing microlensing surveys greatly extended
our knowledge of variable stars.  As the observations of these
experiments usually cover a long time span with a good time sampling
the resulting datasets are suited perfectly for the study of many
different types of variable sources. Lots of the progress in this
field therefore resulted from the work of collaborations like MACHO
(e.g.,~\citet{alcock95,alcock98}), EROS
(e.g.,~\citet{beaulieu95,derue02}), OGLE
(e.g.,~\citet{cieslinski03,wray04}) or MOA
(e.g.,~\citet{noda02,noda04}).  Numerous publications enlarged the
list of known variable stars like Cepheids, RR Lyrae or galactic Long
Period Variables (LPVs), but also helped to understand the physical
processes dominating these stars.

M31 was surveyed for variable sources since the 1920s starting with
the pioneering work of Edwin Hubble. With plates taken at the newly
available Mount Wilson telescope Hubble succeeded to resolve Cepheid
variables in the outer parts of M31.  Using the already established
period-luminosity relation, \citet{hubble} was able to determine the
distance to M31 to 300 kpc (the difference to the actual value of 780
kpc \citep{stanek98} is mainly due to an erroneous calibration of the
zero-point of the PL-relation and a missing reddening correction) and
in this way to reveal the extragalactic nature of the `Andromeda
Nebula'.

\citet{baade63,baade65} continued the work on variable stars in M31
and detected over 400 variables, among them Cepheids and novae.
Looking at the relations between period, luminosity, amplitude and
frequency they found that the Cepheids resemble the ones in the Milky
Way, but seem to be different from those in the Small Magellanic
Cloud.  Another comprehensive study was performed by the DIRECT
project \citep{kaluzny98} who detected and examined Cepheids and
Detached Eclipsing Binaries in five fields towards the M31 disk in order to
reduce the uncertainties in the distance determination of M31.
Unfortunately, the data of a 6th field towards the bulge of M31, which
in part overlaps with our surveyed area, remained unpublished.

In the last years Andromeda was target of several microlensing surveys
(AGAPE \citep{ansari97}, POINT-AGAPE \citep{auriere01}, WeCAPP
\citep{wecapp01}, MEGA \citep{dejong04}, SLOTT/AGAPE
\citep{calchinovati02}) looking for compact dark matter in the halo of
M31. As by-product of these surveys, catalogues of variable sources
start to appear. The work presented in this paper overlaps with the
work resulting from the AGAPE \citep{ansari04} and POINT-AGAPE
\citep{an04} data sets.

Due to our daily stacking scheme of the observations, we are not
sensitive to variations and periods smaller than 1.3 days and focus
therefore on variations of longer period variables, namely Cepheids,
RV Tauri stars, and Long Period Variables (LPVs), i.e. Miras and
semi-regular variables. The catalogue is completed by eclipsing
binaries, and variables showing eruptive or irregular
variations. $\delta$ Scuti and RR Lyrae stars show variation
amplitudes below our detection limit and will therefore be missed in
this study.

The paper is organized as follows. In Sec.~\ref{sec.data} we give an
overview of the survey, the observations, and data
reduction. Sec.~\ref{sec.detection} deals with the source detection
and the derivation of possible periods of the variables. In
Secs. \ref{sec.variables} and \ref{sec.classes}, we show the different
groups of variable sources detected in the survey. The catalogue of
variable stars is presented in Sec.~\ref{sec.catalogue} and is
correlated with X-ray selected catalogues in
Sec.~\ref{sec.comparison}. Whereas Sec.~\ref{sec.conclusions}
summarizes the paper, the appendix deals with the accuracy of the
derived periods.
\section{The dataset \label{sec.data}}
\subsection{Description of the survey}
\begin{figure}[t]
\centering
\includegraphics[width=0.49\textwidth]{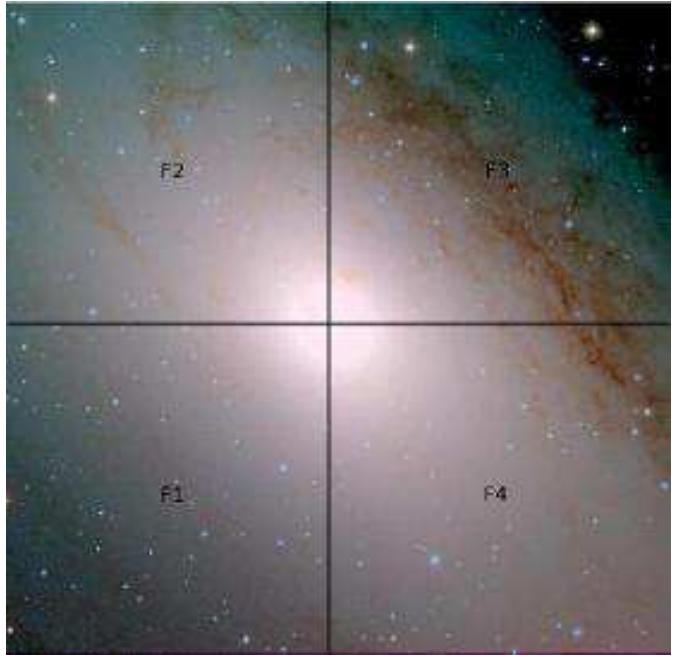}
\caption{\label{fig.m31} $V$-,$R$-, and $I$-band composite image of
the observed fields F1 to F4, taken at Calar Alto Observatory during
the campaign 2000/2001. The black lines mark the positions of fields
F1 to F4. The identified variable sources lie
within RA(2000): [00h43m25.0s, 00h41m59.9s] and DEC(2000):
[41d08$\arcmin$00.1$\arcsec$, 41d24$\arcmin$18.0$\arcsec$].  }
\end{figure}

\begin{figure}[b]
\centering
\includegraphics[width=0.49\textwidth]{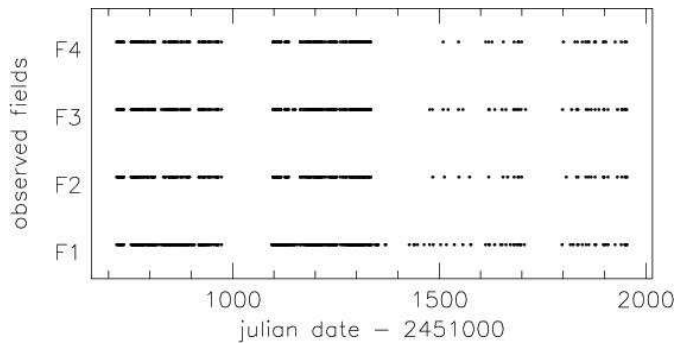}
\caption{\label{fig.sampling} Distribution of the observations for the
four fields.  During the first two campaigns we reached a very dense
time sampling of the observations.  Note that the third and fourth
campaign were restricted to Wendelstein Observatory only.}
\end{figure}

The data presented here result from three years observations of the
central part of M31 by the WeCAPP project. We obtained data from the
0.8~m telescope at Wendelstein Observatory (Germany), and from the
1.23~m telescope at Calar Alto Observatory (Spain).  At Wendelstein
with its field of view (FOV) of $8.3\arcmin \times 8.3\arcmin$ we
selected a field (F1 in the following) along the minor axis of M31
which contains the area with the highest expected rate for
pixellensing events (see also Fig.~\ref{fig.m31} for the location of
F1). Observations of this field were accompanied by images of F3, the
opposite field along the NW minor axis, taken with a sparser time
sampling.  The Calar Alto field covered $17.2\arcmin \times
17.2\arcmin$ centered on the nucleus of M31. Two quadrants of the
field coincided with the maximal lensing field F1 and the opposite
field F3. Due to the simultaneous observations we reached a very good
time sampling during the observability of M31. Since summer 2002 we
are mosaicing fields F1 to F4 with the Wendelstein telescope solely. A
composite image ($V$-, $R$-, and $I$-band) of fields F1 to F4 taken at
Calar Alto Observatory during the campaign 2000/2001 is shown in
Fig. \ref{fig.m31}. The epochs with data taken for the four fields are
shown in Fig. \ref{fig.sampling}.

\subsection{Instruments and filters}  
An overview of the cameras and CCDs used in WeCAPP is shown in
Table~\ref{tab.ccd}. The cameras are a 1K~$\times$~1K TEK~camera with
a pixel size of $0.49\arcsec$ at Wendelstein and a 2K~$\times$~2K SITe
camera with a pixel size of $0.5\arcsec$ at Calar Alto. The
observations were carried out in $R$ and $I$ filters close to the
Kron-Cousins system.  At Wendelstein we used the $R$
($\lambda\simeq650\,\mathrm{nm}$,
$\Delta\lambda\simeq150\,\mathrm{nm}$) and $I$
($\lambda\simeq850\,\mathrm{nm}$,
$\Delta\lambda\simeq150\,\mathrm{nm}$) wavebands. The Calar Alto
observations were carried out with the very similar filters,
$R$($\lambda\simeq640\,\mathrm{nm}$,
$\Delta\lambda\simeq160\,\mathrm{nm}$) and $I$
($\lambda\simeq850\,\mathrm{nm}$,
$\Delta\lambda\simeq150\,\mathrm{nm}$).  Despite of the combination of
different telescopes, CCDs, and slightly different filter systems we
observed no systematic effects in the light curves depending on these
parameters (see also Fig.~\ref{fig.reg_lpv} where we show the data
obtained at Calar Alto and at Wendelstein separately.)
\subsection{Data reduction}
The WeCAPP reduction pipeline {\sf mupipe} will be presented in a
future publication \citep{mupipe}. {\sf mupipe} combines all reduction
steps from de-biasing of the images until PSF (point-spread-function)
photometry and the final measurements of the light-curve in one
software package, including full error propagation from the first
reduction step to the last \citep{redux}:
\begin{enumerate}
\item standard CCD reduction including de-biasing, flatfielding and
filtering of cosmic ray events
\item position alignment on a reference grid using a flux conserving 
interpolation routine
\item photometric alignment of the images
\item restoration of pixels damaged by CCD defects (cold and hot
pixels, traps, bad columns) or pixels hit by cosmic ray events
\item stacking of the frames of one epoch (i.e. night) using a weighting 
scheme to maximize the $S/N$ for point sources
\item matching of the PSF of a high $S/N$ reference frame to the PSF
of each stacked frame using our implementation of the image
subtraction method developed by \citet{alard98} and \citet{alard00}
\item generation of difference images by subtracting the convolved
reference frame from each stacked frame
\item PSF photometry of each pixel in the difference frames using a PSF
extracted from the convolved high $S/N$ reference frame
\item extraction of the light curves
\end{enumerate}

\begin{table*}[t]
\centering
\begin{tabular}{llcccccc}
\hline \hline Site & Campaign & CCD  & Size &
[$\mathrm{arcsec}/\mathrm{px}$] & Field [$\mathrm{arcmin}^2$]\\ \hline
We & 2000-2003 & TEK \#1      & 1K $\times$ 1K & 0.49 & $8.3   \times 8.3$  \\
CA & 2000-2002 & SITe2b \#17  & 2K $\times$ 2K & 0.50 & $17.2  \times  17.2$  \\ 
\hline\\
\end{tabular}
\caption{Properties of the CCD cameras used during WeCAPP at
Wendelstein (We) and Calar Alto (CA) Observatories, respectively. Both
CCDs have a pixel size of $24\,\mu\mathrm{m}$.}
\label{tab.ccd}
\end{table*}

\begin{figure}[th]
\centering
\includegraphics[width=0.49\textwidth]{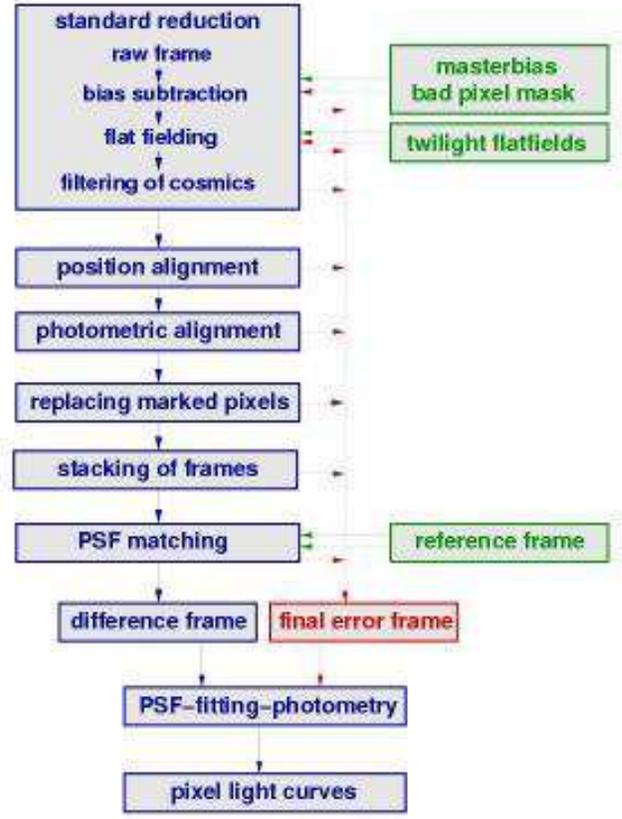}
\caption{Floating chart of the WeCAPP reduction pipeline {\sf mupipe}.
The reduction package includes full error propagation for each pixel
through all reduction steps. In this way, all data points are properly
taken into account in the search for variable sources.  The reduction
pipeline returns light curves for each pixel in the frame,
representing the temporal change of the flux present inside the PSF
centered on the particular pixel. \label{fig.mupipe}}
\end{figure}

\noindent A floating chart of the reduction pipeline is shown in
Fig.~\ref{fig.mupipe}. In the last step {\sf mupipe} returns roughly 4
$\times 10^6$ pixel light curves together with appropriate errors,
each of the light curves representing the time variability of the flux
present inside the PSF centered on the particular pixel.  The
extraction of intrinsic variable sources from these pixel light curves
is presented in the next section.

\subsection{Astrometry\label{sec.astrometry}}

\begin{figure}[t]
\begin{center}
\includegraphics[width=0.4\textwidth,angle=0]{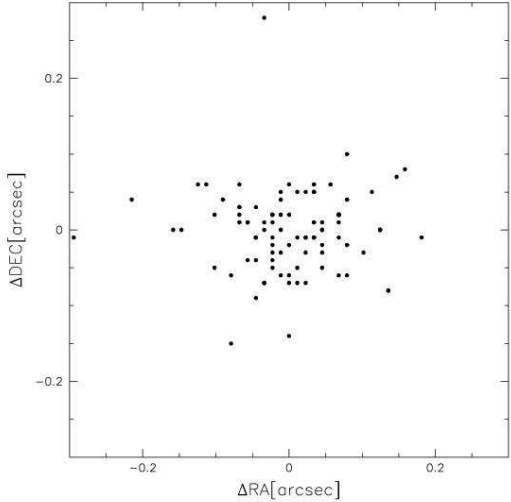}
\end{center}
\caption{\label{fig.astro_lgs} Errors of the astrometric solution for
the calibration stars taken from the LGS survey. The rms is
0.10$\arcsec$ in declination and 0.12$\arcsec$ in right ascension.}
\end{figure}

The astrometric solution was created using the IRAF\footnote{IRAF is
distributed by the National Optical Astronomy Observatories, which are
operated by the Association of Universities for Research in Astronomy,
Inc., under cooperative agreement with the National Science
Foundation.} tasks {\sf ccmap} and {\sf cctran} for 92 stars whose
positions were taken from the Local Group Survey (LGS, \citet{lgs}).
The rms of the solution is 0.10$\arcsec$ in declination and
0.12$\arcsec$ in right ascension. The errors of the astrometric
solution for the 92 calibration stars are shown in
Fig.~\ref{fig.astro_lgs}.  The coordinates derived here agree
perfectly with the one's derived in \citet{wecapp01} and
\citet{wecapp03}.

%
\section{Detection of variable sources \label{sec.detection}}

\subsection{Selection of the sources}

\begin{figure}[b]
\centering
\includegraphics[width=0.4\textwidth]{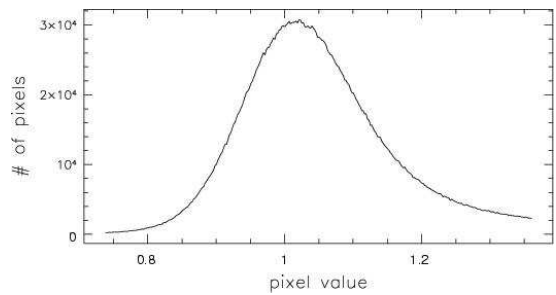}
\caption{\label{fig.chi2hist} Histogram of the pixel values measured
in the $\chi^2_\nu$-frame. The maximum of the distribution of 1.02
reflects the accuracy of the propagated errors.}
\end{figure}

For the selection of the intrinsic variable sources we use the
$R$-band data from the Calar Alto campaigns 2000/2001 and 2001/2002 to
create a $\chi^2_\nu$ variation frame. To get rid of systematic
effects induced by the different seeing conditions of each frame we
use the following approach. We select a stacked image with roughly the
median seeing of the 2000/2001 campaign (about 1.5$\arcsec$) as master
frame and match the PSF of each stacked frame with a smaller PSF to
the one of the master frame.  In this way we end with difference
images for 113 epochs (i.e. 48~\% of the epochs with $R$-band data
obtained at Calar Alto) which are used for the detection of the
variable sources. For each of the light curves extracted from the
PSF$_{1.5}$ difference images we calculate the reduced $\chi^2_\nu$
deviation from a constant (i.e. zero) baseline fit.  The errors
entering this calculations are the propagated errors as returned from
{\sf mupipe}. The results are written into a $\chi^2_\nu$-frame of the
field. The mode of the $\chi^2_\nu$-frame is 1.02 which shows the
accuracy of the propagated errors (see Fig. \ref{fig.chi2hist}). Each
value $\ge$ 1.15 is connected to a non-constant, i.e. variable source
at the 99.99~\% confidence level.

To avoid contamination of the M31 sample with foreground objects we
set all pixels in a radius of 5 pixels around the positions of bright
foreground stars to zero before detecting the variable sources in the
$\chi^2_\nu$-frame.
\begin{figure}
\centering
\includegraphics[width=0.4\textwidth]{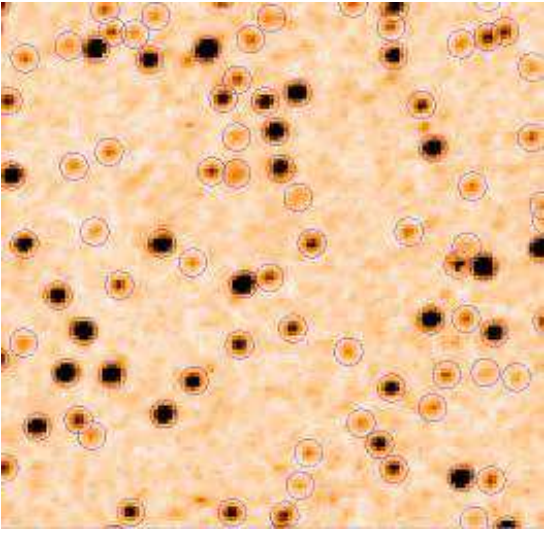}
\caption{\label{fig.chi2} Extract ($1\arcmin \times 1\arcmin$) of the
$\chi^2_\nu$-frame of the over 25000 variable sources detected by the
WeCAPP microlensing survey of M31. The circles give the positions of
the sources detected by SExtractor.  Most of the sources are first
identifications, the majority of them being Long Period Variables
(LPVs), i.e. Miras and semi-regular variables (see
Sec. \ref{sec.variables}).}
\end{figure}
We determine the positions of the variable sources using the
SExtractor software \citep{sextractor} for source detection applied on
the $\chi^2_\nu$-frame. In this way we detect 25571 variable sources
in our observed field. Figure \ref{fig.chi2} shows an extract of the
$\chi^2_\nu$-frame with the positions of the detected sources marked.
\subsection{Period determination}
We use an algorithm developed by \citet{lomb} and \citet{scargle} to
determine the significance and value of a possible period of the
variables sources. The method of Lomb is well suited for this problem
as it is able to deal with unevenly sampled data.

\begin{figure}[t]
\centering
\includegraphics[width=0.49\textwidth,angle=0]{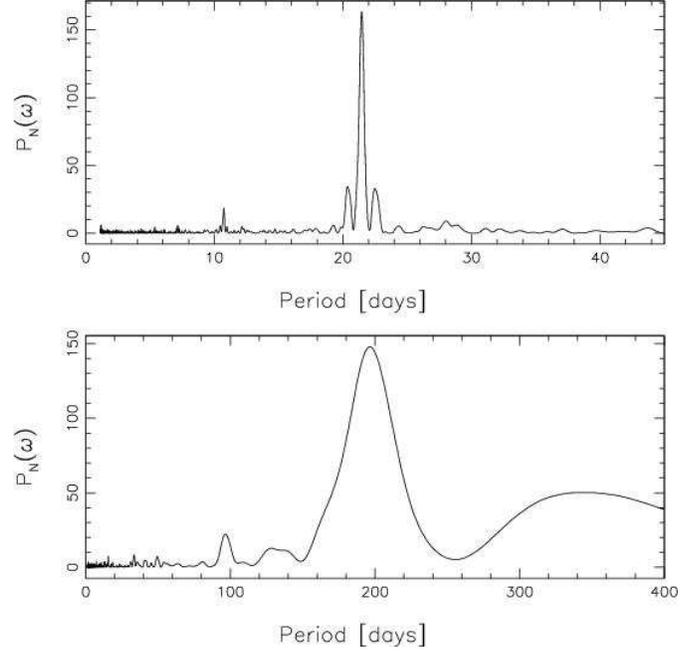}
\caption{\label{fig.spectrum} Two examples of power spectra
$P_N(\omega)$ of WeCAPP variables. In the upper panel we show the
periodogram for a Cepheid (see also Fig.~\ref{fig.d_ceph}) derived
from the $R$-band data, in the lower panel we show the periodogram for
a Long Periodic Variable (see also Fig.~\ref{fig.reg_lpv}) derived
from the $I$-band data.}
\end{figure}

The algorithm extracts the power in the first sine and cosine terms
for a set of equidistantly spaced frequencies and also yields the
significance of the detected peaks in the power spectrum. The Lomb
normalized periodogram for $N$ measured data points $h_j=h(t_j)$,
$j=1,...,N$, taken at epochs $t_j$ is defined by

\begin{eqnarray}
P_N(\omega) & = &
\frac{1}{2\sigma^2} \frac{\left[\sum_j(h_j-\bar{h})\cos \omega \,(t_j-\tau)\right]^2 }{\sum_j\cos^2 \omega\,(t_j-\tau)}\nonumber\\
 && +\frac{1}{2\sigma^2} \frac{\left[\sum_j(h_j-\bar{h})\sin\omega\,(t_j-\tau)\right]^2}{\sum_j\sin^2 \omega \,(t_j-\tau)}\quad,
\label{eq.pn}
\end{eqnarray}

where $\omega=2\pi \nu= \frac{2\pi}{P}$ is the angular frequency for
the period~$P$, and the mean $\bar{h}$, the variance $\sigma^2$, and
the constant $\tau$ are defined as follows

\begin{equation}
\bar{h}\equiv\frac{1}{N}\sum_{j=1}^N h_j \,\,\,,
\label{eq.hm}
\end{equation}
\begin{equation}
\sigma^2\equiv\frac{1}{N-1}\sum_{j=1}^N (h_j-\bar{h})^2 \,\,\,,
\label{eq.hsig}
\end{equation}
\begin{equation}
\tan(2\omega\tau)\equiv\frac{\sum_j \sin 2\omega t_j}{\sum_j \cos 2\omega t_j} \,\,\,.
\label{eq.tau}
\end{equation}
The last definition ensures the power spectrum to be independent on a
shift of each $t_j$ by $\Delta t$, as $\tau$ transforms in that case
into $(\tau + \Delta t)$. \citet{lomb} has shown that the evaluation
of the periodogram according to Eq. \ref{eq.pn} is identical to a
linear least-squares fit of the first harmonics

\begin{equation}
F(t)=A \sin \omega t + B \cos \omega t
\end{equation}

to the data points. In Fig.~\ref{fig.spectrum} we show two examples
for $P_N(\omega)$, the first one derived from a Cepheid light curve,
the second one derived from the light curve of a Long Periodic
Variable.

The significance level $P(>z)$ of a peak with amplitude 
$z$ in the power spectrum is given by

\begin{equation}
P(>z)\equiv 1 - (1-e^{-z})^{M} \quad,
\end{equation}

with $(1-e^{-z})^{M}$ being the probability that none of the $M$
tested frequencies shows an amplitude greater than $z$ in case of pure
Gaussian noise.  \citet{horne} showed with extensive Monte Carlo
simulations, that the value of $M$, i.e. the number of independent
frequencies does not differ much from the number of data points $N$,
if the data points are not closely clumped.

We use the implementation of the algorithm taken from the Numerical
Recipes \citep{press}. Small modifications in the code allow us to
search for the different maxima in the power spectrum. The positions
and values of the peaks are returned and can be used for further study
of the light curves. In this implementation the significance $P(>z)$
is approximated for small values of $P(>z)$ by

\begin{equation}
P(>z)\approx M e^{-z} \quad,
\end{equation}

where $M$ is a product of the number of data points $N$ and a user
supplied value, which determines the high end cut-off of the tested
frequencies relative to the Nyquist frequency.
\section{The variable sources \label{sec.variables}}
\subsection{Creating the catalogue\label{sec.refining}}
We detect 25571 variable point sources in a
16.1\arcmin~$\times$~16.6$\arcmin$ field centered on the nucleus of
M31.  As this sample still can contain spurious detections or sources
with only a few data points measured in the light curve we apply a
couple of cuts which mark different levels of accuracy of the derived
periods and the required need for a visual inspection.

\begin{figure}[t]
\centering
\includegraphics[width=0.4\textwidth]{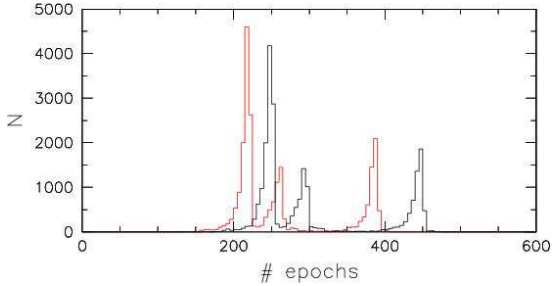}
\caption{\label{fig.hist_epochs} Histogram of the number of epochs of
the level A sources in the $R-$ band (black curve) and $I$-band (red
curve). For sources located in field F1 we usually got data points in
the $R$-band on more than 400 epochs.}
\end{figure}

As first cut we demand that the light curve comprises at least 40 data
points in both filters. All sources which do not pass this criterion
are removed from the sample.  This reduces the number of variables
under inspection to 25316 and defines the {\bf level~A sample}.  In
Fig. \ref{fig.hist_epochs} we show the histogram of the number of
epochs for these sources in the $R$-band (black curve) and $I$-band
(red curve). The lowest number peaks represent the time sampling
present in fields F2 and F4, followed by the peak corresponding to
field F3. As F2 and F4 were observed with comparable frequency both
fields contribute to the lowest number peak. The last peak with
usually more than 400 epochs in the $R$-band is related to field F1.

As the Lomb algorithm assumes that all data points have the same error
(see Eqs. \ref{eq.pn} and \ref{eq.hsig}), data points with big error
bars or outliers can spoil the period finding process and yield
spurious periods.  To avoid this we first eliminate in each light
curve ten data points with the biggest errors and in a second step
five data points with the highest and lowest values, respectively. In
doing so, we ensure that we have at least 20 data points in both
filters to look for a periodic signal.

We now check the $R$-band and $I$-band light curves separately for
periodicities. As the color of variable stars usually changes during a
cycle combing the two data sets would not result in accurate
determinations of the period of the sources.  Finally we obtain 25316
variables with determined periods (of any significance level) in both
bands.

We regard those periods as real which are the same in both bands
inside tight boundaries reflecting the error of the period
determination.  For periods $P_R <215$ days we take the theoretical
error (see Eq.~\ref{eq.p_error}) resulting from $\nu_{FWHM}$ as limit,
from 215 days onwards we choose a more conservative constant limit of
30 days, which is increased to 60 days for $P_R>400$ days.  This
criterion defines the {\bf level~B sample} (20311 objects).

As final cut we select all variables from level B which show a
significance level of the period determination of $P(>z)$ $<$
$10^{-10}$ in $R$ or $I$ and which have a determined $R$-band period
$P_R\le450$ days.  This final cut reduces the number of sources to
19551 and defines the {\bf level~C~sample} with well determined
periods. The remaining sources in level B were inspected visually and
assigned to the group of regular, irregular, long-variation (period
could not be determined because of an incomplete cycle) or
miscellaneous variables in the {\bf final catalogue}, or rejected from
the sample as spurious detections.  Finally, all sources of the level
C sample are added automatically to the final catalogue.

For the variables which show deviant periods in $R$ and $I$ we proceed
as follows.  If the significance level of one of the deviant periods
is better than $10^{-15}$ and at the same time better than the
significance in the other band by a factor of $10^{10}$ we choose the
period in the first band and add the variable to the final catalogue.
If this is not the case we inspect the light curve visually and decide
if one of the periods is the real one, or if the variation is of
irregular nature. In this step we also reject the last spurious detections
from the sample, the true variables are added to the final catalogue.
\subsubsection{Search for eclipsing binaries}
By visual inspection of the folded light curves of the Lomb sample we
detect 28 eclipsing binary (EB) candidates, amongst them one
(semi-)detached system.  However, the Lomb algorithm can fail in
detecting periods for potential EB systems in M31. As EB light curves,
especially the ones of detached systems, show strong power in the
higher harmonics, they are not well recovered by Fourier techniques
using the first harmonic only. Since some EB candidates therefore
could be missed by using the Lomb algorithm alone, we add another step
to the search for EBs.  We investigate all light curves which were
removed in the previous steps once more, however use this time the
dedicated transit finding algorithm `{\sf boxfitting}' \citep{kovacs}
for the period determination. As \cite{tingley} has shown the
boxfitting algorithm is a powerful tool for the detection of eclipsing
systems. We implemented the original Fortran
code\footnote{http://www.konkoly.hu/staff/kovacs.html} into our
detection pipeline and run the algorithm over the mostly noisy light
curves which were removed from the sample as spurious detections in
the previous steps. In this way we assure to have proper periods for
all potential EB candidates present among our variable sources.

\begin{figure}[t!]
\centering
\includegraphics[width=0.46\textwidth]{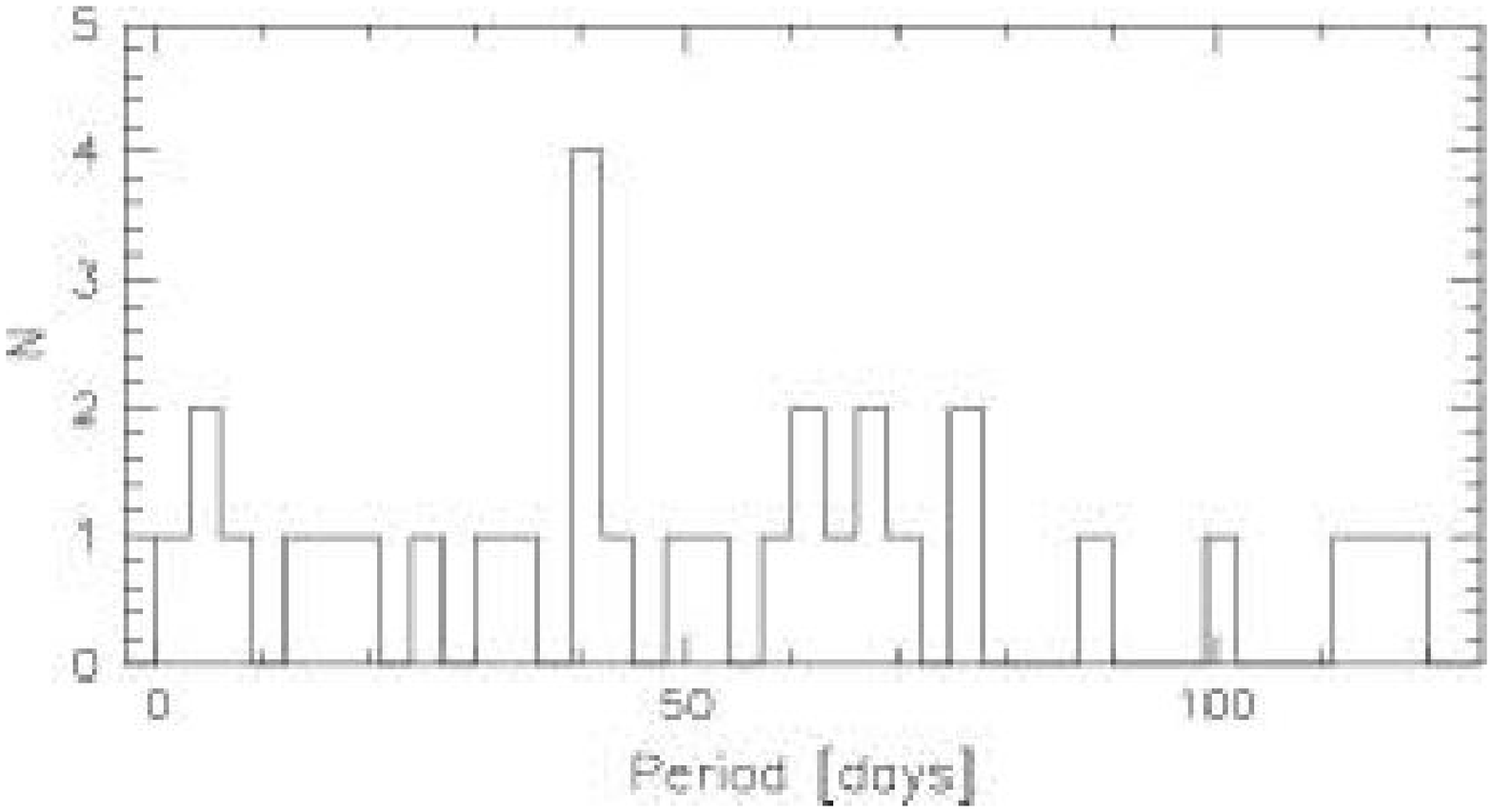}
\hfill
\includegraphics[width=0.46\textwidth]{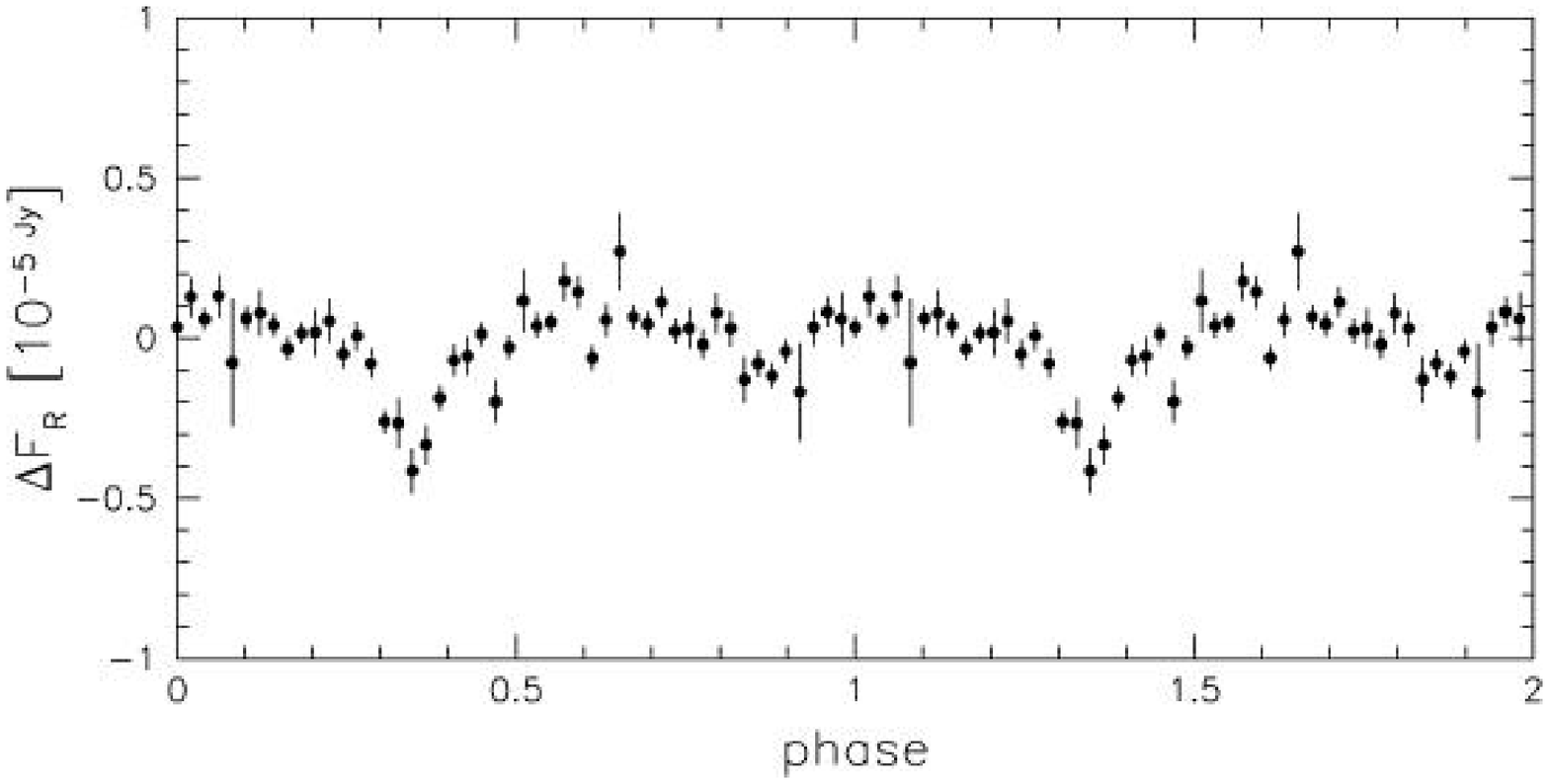}
 \caption{\label{fig.EB.phist} Top panel: Histogram of the periods of
the 31 eclipsing binary candidates. The 4 (semi-)detached systems
populate the low period area up to a period of 7 days whereas the
candidate contact systems generally have longer periods. Bottom panel:
binned $R$-band light curve of a semi-detached system in the phase
representation. The system has a $R$-band period of 7.08 days.  }
\end{figure}

Selecting all light curves with a reasonable signal detection
efficiency (SDE, \citet{kovacs}) SDE $>6$, periods $P>1.30$ days, and
rejecting periods between 1.95 days and 2.05 days (to avoid to pick up
aliasing periods) results in an additional sample of 155 light
curves. Visual inspection of the folded light curves finally yields 3
additional candidates for (semi-)detached eclipsing binary systems.

In Fig.~\ref{fig.EB.phist} we show the distribution of periods of the
31 EB candidates in the upper panel. The 4 (semi-)detached systems
populate the low period area up to a period of 7 days whereas the
candidate contact systems generally have longer periods. In the bottom
panel we show the binned $R$-band light curve of a semi-detached
system in the phase representation. The system was detected using the
Lomb algorithm and has a $R$-band period of 7.08 days.
\subsection{Number counts and asymmetry}
\begin{figure}[t!]
\centering
\includegraphics[width=0.49\textwidth]{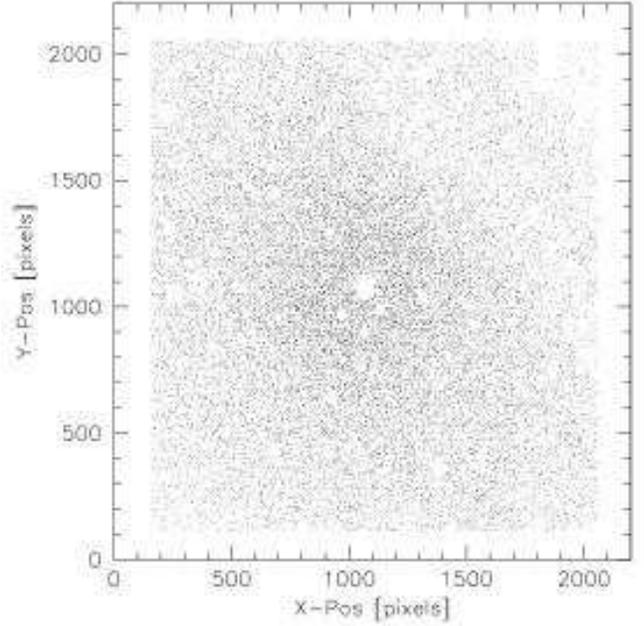}
\caption{\label{fig.positions} Positions of the 23781 variable sources
of the final catalogue.  Blank regions are connected to to saturated
parts of the frame or to eliminated foreground stars.  Incompleteness
induced by crowding in the center and the influence of the enhanced
extinction connected with the spiral arms superimposed over the M31
bulge are clearly visible (see also Fig.~\ref{fig.numdens}).}
\end{figure}

\begin{figure}[h!]
\centering
\includegraphics[width=0.4\textwidth,angle=0]{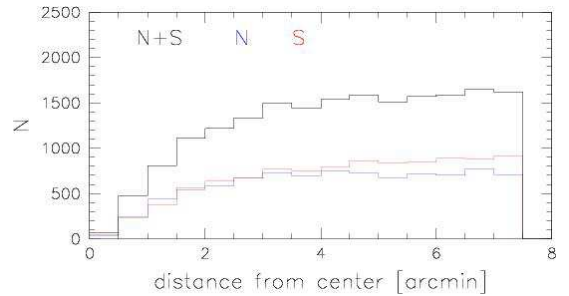}
\caption{\label{fig.nc} Number counts of sources of the catalogue as
function of the distance to the center of M31.  Black: all
sources. Blue: sources in the northern hemisphere of M31.  Red:
sources in the southern hemisphere of M31. Clearly visible is the
asymmetry of the detected sources due to the enhanced disk-extinction
in the northern part. The low number of sources near the center is due
to incompleteness induced by crowding and saturated parts in the
frames.}
\end{figure}

In Fig. \ref{fig.positions} we show the positions of the variables
from the final catalogue, which suggests a connection of the enhanced
extinction in M31's spiral arms and the depletion of sources in
certain regions in the northern part of the bulge (field F3). This
depletion is also evident from number counts of the sources in the
northern and southern hemisphere of M31 (see Fig.\ref{fig.nc}).

We are calculating the number density of variable sources by summing
up all variables in 100$\times$100 pixel [$50\arcsec \times
50\arcsec$] bins.  In the resulting density map
(Fig.~\ref{fig.numdens}) the spiral arms are clearly visible. A
comparison of the number densities at equivalent positions in the M31
bulge, one in the dust lane of the spiral arms, the other one in the
opposite hemisphere, shows a reduction to about 60\% relative to the
part with no strong extinction.

To examine this subject further we compare the positions map (see
Fig. \ref{fig.positions}) with an extinction map of our field.  In
Fig.~\ref{fig.ext} we plot the positions of the sources on top of the
$R$-band extinction map, which we derived from the $V$- and $R$-band
frames taken by the LGS survey in the following way:

\begin{figure}[t]
\begin{center}
\includegraphics[width=0.4\textwidth,angle=0]{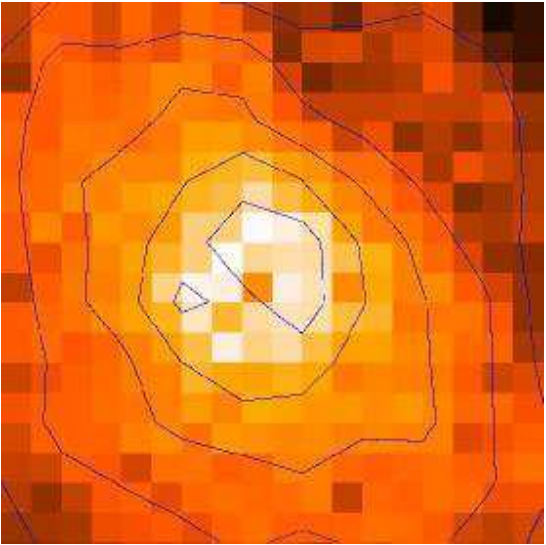}
\caption{\label{fig.numdens} Number density of variable sources by
summing up all variables in 100$\times$100 pixel [$50\arcsec \times
50\arcsec$] bins. To avoid effects induced by the border of the frame
we only use pixels in the region [200:2000,200:2000]. The contour
levels are 1, 2, 3, 4 and 5 $\times 10^{-2}$ [arcsec$^{-2}$],
respectively. The spiral arm is clearly visible. Due to extinction in
the dust lanes the number density reduces to about 60\% when compared
to equivalent regions without strong dust features being present. The
regions with underdensities near the center are due to saturated parts
in the frame with no variable sources detected.}
\end{center}
\end{figure}

We start from the relation between the color excess $E(V-R)$ and the
extinction $A_R$ in the $R$-band ($a \approx 3$, see
\citet{binney_merrifield})
\begin{equation}
A_R = a \,\,E(V-R)
\end{equation}
With the non-reddened magnitudes $M_{R,0}, M_{V,0}$ and reddened counterparts
$M_{R,r}, M_{V,r}$ this writes as
\begin{equation}
M_{R,r}-M_{R,0}= a \,\, [(M_{V,r}-M_{R,r})-(M_{V,0} - M_{R,0})] \,\,\,.
\end{equation}
The transformation to fluxes $F_{i,j}$ ($i=R,V; j=0,r$) yields
\begin{equation}
F_{R,0}=F_{R,r} \left(\frac{F_{R,r}}{F_{V,r}}\right)^a \left(\frac{F_{V,0}}{F_{R,0}}\right)^a \,\,\,.
\end{equation}
If we now suppose that the intrinsic stellar population gradients are
negligible over the field we can set 
$\left(\frac{F_{V,0}}{F_{R,0}}\right)\approx~constant$, which is not
exactly true but a valid approximation for our purposes.

\begin{figure}[t]
\begin{center}
\includegraphics[width=0.418\textwidth,angle=0]{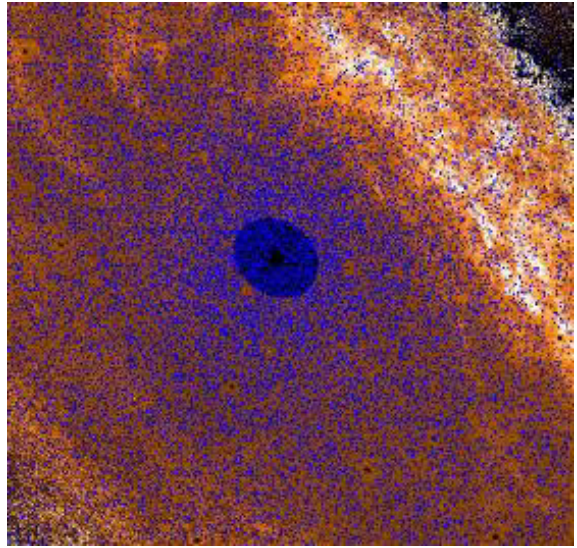}
\caption{\label{fig.ext} $R$-band extinction map of the M31 field.
White areas correspond to high extinction values. The extinction in
the $R$-band was derived using the $V$- and $R$-band images taken from
the LGS survey \citep{lgs}. Overplotted are the positions of the
catalogue variables (blue dots). In the central regions the extinction
map could not be calculated due to saturation in the original LGS
frames.}
\end{center}
\end{figure}

Using this assumption
we finally obtain a relation for $F_{R,0}$
\begin{equation}
F_{R,0}=\frac{F_{R,r}^{1+a}}{F_{V,r}^a}  \,\,\, \left(\frac{F_{V,0}}{F_{R,0}}\right)^a \,\,\,,
\end{equation}
which can be used to calculate the extinction $ext_R$ in the $R$-band
\begin{equation}
ext_R = -2.5 \log\left(\frac{F_{R,r}}{F_{R,0}}\right) \,\,\,.
\end{equation} 

 For the intrinsic color $(V-R)_0$ required in the calculation we use
the theoretical value $(V-R)_0=0.63$, assuming the bulge to be a 12
Gyr old SSP of 2$Z_\odot$ metalicity (C. Maraston, priv. comm.).
Figure~\ref{fig.ext} shows the resulting extinction map: indeed
under-dense variable source regions coincide with high extinction
regions.

\begin{figure}[t]
\centering
\includegraphics[width=0.4\textwidth,angle=0]{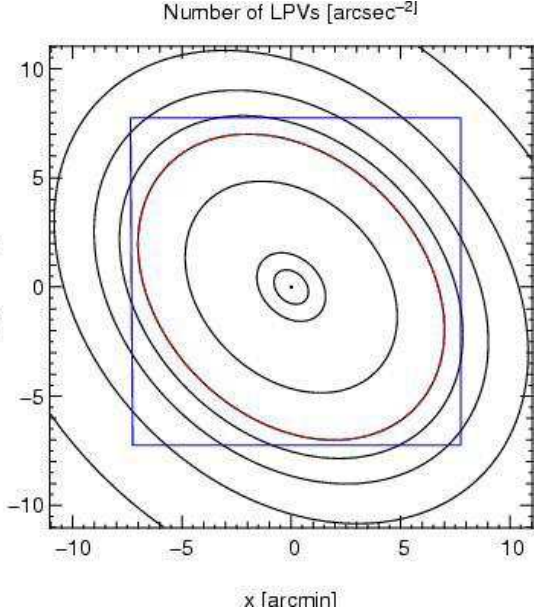}
\caption{\label{fig.fuel} Theoretical expected number densities of
LPVs according to \citet{renzini98} calculated with the small bulge
decomposition of \citet{kent89} . The field is centered on the nucleus
of M31, north is on top, east is on the left of the diagram.  The
WeCAPP field for which we calculated the number density in
Fig.~\ref{fig.numdens} is shown by blue lines. Contour levels are 1,
2, 3, 4 and 5~$\times 10^{-2}$, 0.1, 0.5, 1, 5 and 10 sources
[arcsec$^{-2}$]. The red line marks the 5~$\times 10^{-2}$ contour
level.}
\end{figure}

Finally we compare the number densities of sources present in our
catalogue with theoretical predictions. According to \citet{renzini98}
the number of LPVs per integrated $10^5$ bolometric luminosities is
equal to 0.5. \citet{renzini98} uses a slightly older (15 Gyrs) and
less metal-rich (1 $Z_\odot$) bulge if compared to the model used for
the calculation of the extinction frame. With the small bulge
decomposition of \citet{kent89} and under the assumption that the LPVs
are present in the bulge only, we derive the expected number densities
of LPVs and show it in Fig.~\ref{fig.fuel}.  The agreement with the
detected number densities is good in the outer parts of the
field. Towards the center we suffer from incompleteness due to
enhanced noise on the one hand and crowding of the sources on the
other hand.
\subsection{Classification scheme}
As practically all of the detected sources are unresolved in the
original frames we are neither able to put the sources in the
color-magnitude plane and to construct a color-magnitude diagram, nor
to derive period-luminosity relations for the different classes. The
available parameters for establishing a classification scheme are
therefore reduced, leaving the period, its significance, the amplitude
of the variation, its (flux excess)-color, and finally the light curve
shape as classification parameters. The light curve shape is of
particular interest as it can be described mathematically and
parameterized in terms of the parameters resulting from low order
Fourier fits to the data. For classical Cepheids the Fourier
parameters show a progression with the period of the variation,
echoing the well-known `Hertzsprung progression' \citep{hertzsprung}
of the light curve shape with period. For type II Cepheids likewise
correlations have already been found; the analysis of the RV Tauri
stars in our sample reveals a correlation between different phase
parameters.  Fourier decomposition of the light curves therefore is a
powerful method to support the classification of Cepheid-like
variables and discriminate them from other type of variables.  On
first ground our classification scheme is based on the position of the
stars in the $R$-band period-amplitude plane. For certain groups of
stars we refine and check our classification by using the Fourier
parameters of the fits to the light curves.
\subsubsection{Period-amplitude-relations}
The amplitudes $\Delta F_R$ and $\Delta F_I$ of the variation in the
$R$-band and $I$-band are determined as half the difference of the
maximum to the minimum of the light curves.  As we calculate the
amplitudes after having eliminated data points with the largest error
bars and biggest flux differences this amplitudes act as a lower limit
only, but can be regarded as a robust measurement of the variation
amplitude.  The amplitudes were transformed in magnitudes using the
$R$- and $I$-band fluxes of Vega taken from
\citet{binney_merrifield}. Note that these magnitudes reflect the flux
difference on the frame only, and are not the real variation
magnitudes of a single star.

\begin{figure}[t]
\centering
\includegraphics[width=0.4\textwidth,angle=0]{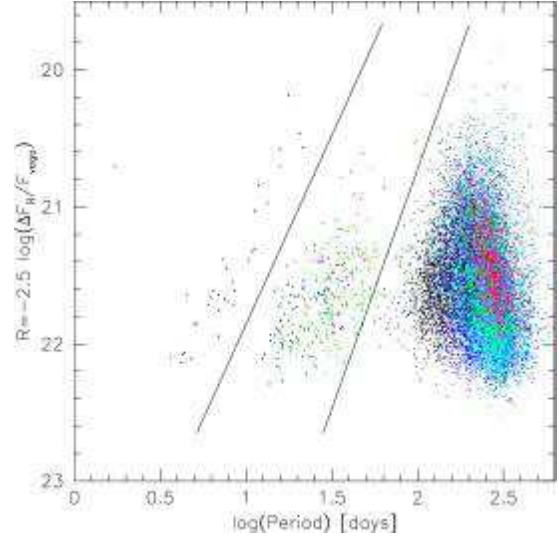}
\caption{\label{fig.peramp} Period-amplitude relation in the $R$-band
for the catalogue sources (eclipsing binaries, irregular,
long-variation, and miscellaneous variables excluded).  The amplitudes
were transformed in magnitudes using the $R$-band fluxes of Vega. Note
that these magnitudes reflect the flux difference only and are not the
real variation magnitude of a single star. In this diagram three
populations are visible. The black lines show the defining relations
(from left to right) for group~I (population I Cepheids: black dots),
group~II (type II Cepheids: blue dots, RV Tauri stars: red dots, RV
Tauri candidates: open magenta circles, SR variables: green dots), and
group~III (LPVs). For the RV Tauri stars and candidates we use the
single or fundamental period (minimum to minimum).  The sources in
group III are color coded according to the significance of the
$I$-band period as returned by the Lomb algorithm (low significance:
black $\rightarrow$ blue $\rightarrow$ red: high significance).}
\end{figure}

Figure~\ref{fig.peramp} shows the relation between the logarithm of
the period and the logarithm of the amplitude $\Delta F_R$ (i.e. the
variation magnitude as measured in the frame) for the sources of the
final catalogue.  Three different groups can be distinguished in this
diagram:
\begin{itemize}

\item Group I has periods between 1.7 and 21.5 days, there seems to be
a correlation of the period and the amplitude of the variation in a
sense that bigger periods show larger amplitudes.  We require $-2.5
\,\log(\Delta F_R/F_{\mathrm{Vega}}) < 23.1-2.76 \,(\log P_R -0.55)$
for sources belonging to this group. This relation as well as the
dividing relation for group II sources is shown as black line in
Fig.~\ref{fig.peramp}.  These stars are most likely connected to the
disk, as their light curves and the period regime are connected to
Cepheids of type I which belong to a young population.

\item Group II has periods in the range between 12 and about 140 days.
Also in this group the period and the variation amplitude are
correlated, whereas the sequence for the group II stars lies at
fainter magnitudes than the one of the group~I stars. The defining
relations for this group are given by $-2.5 \,\log(\Delta
F_R/F_{\mathrm{Vega}}) > 23.1-2.76 \,(\log P_R-0.55)$ and $-2.5
\,\log(\Delta F_R/F_{\mathrm{Vega}}) < 23.0-3.5 \,(\log P_R-1.35)$.
Group II stars most likely belong to the old spheroid population; RV
Tauri stars, type II Cepheids and the low period tail of the
semi-regular (SR) stars are found in this group.

\item The bulk of the detected variables finally belongs to group III,
variables with periods longer than about 50 days. For this group no
clear correlation between the amplitude and the period can be seen.
All sources with $-2.5\, \log(\Delta F_R/F_{\mathrm{Vega}}) >
23.0-3.5\, (\log P_R -1.35)$ belong to this group.
\end{itemize}
\subsubsection{Fourier fits \label{sec.fourier}}
We fitted truncated Fourier series of the form

\begin{equation}
C + A_0 \,\Sigma_{i=1}^N \,A_i \cos(i \omega (t-t_0)+\Phi_i)
\end{equation}
to the $R$-band light curves of groups I and II.  $C$ defines the
baseline of the fit, $A_0$ reflects the overall amplitude of the light
curve, and $A_i$ and $\Phi_i$ define the amplitudes and phases of the
different harmonics. The periods entering the fit are the periods
returned by the Lomb algorithm. A fitting order of $N=5$ leads to
acceptable fits to the data.

\cite{simon81} were the first to calculate amplitude ratios of
the form $R_{ij}=A_i/A_j$ and phase differences $\Phi_{ij}=\Phi_i-i\Phi_j$ 
of the parameters of the different harmonics. 
Since for classical Cepheids both definitions show a progression with the 
period as well as a correlation among one another, Fourier analysis 
proved to be an excellent diagnostic to examine the pulsation properties
of these stars. Fourier analysis therefore was widely used in the past 
to examine Cepheids (e.g., \cite{simon86}, \cite{alcock99}), particularly 
to distinguish between fundamental and first overtone pulsators 
(e.g., \cite{antonello}, \cite{beaulieu95}).
\section{Classes of variables \label{sec.classes}}
\subsection{Group I and II - Cepheid-like variables}
Groups I and II are populated by Cepheid-like variables (population I
Cepheids in group I, type II Cepheids and RV Tauri stars in group II),
and the small period tail of semi-regular variable stars, which start
to contribute at periods from about 16 days onwards.
\subsubsection{Population I Cepheids}
Classical Cepheids are relatively young, intermediate mass
population~I stars. They pulsate in the fundamental mode which
discriminates them from s-Cepheids, which are believed to pulsate in
the first overtone. Both groups of population I stars have distinctive
light curves, the classical showing skewed, the s-type showing smooth
sinusoidal variations.

\begin{figure}
\centering
\includegraphics[width=0.49\textwidth,angle=0]{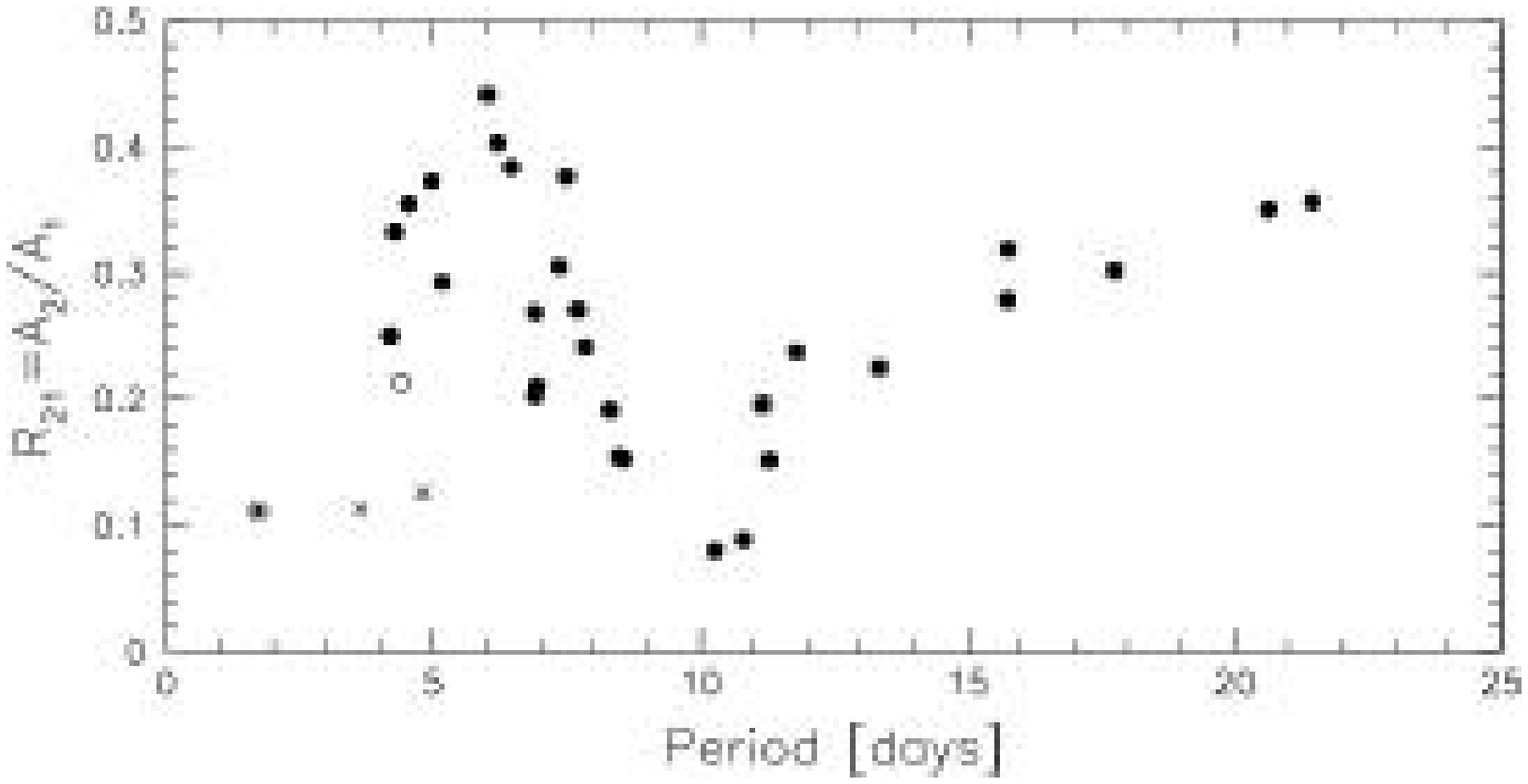}
\hfill
\includegraphics[width=0.49\textwidth,angle=0]{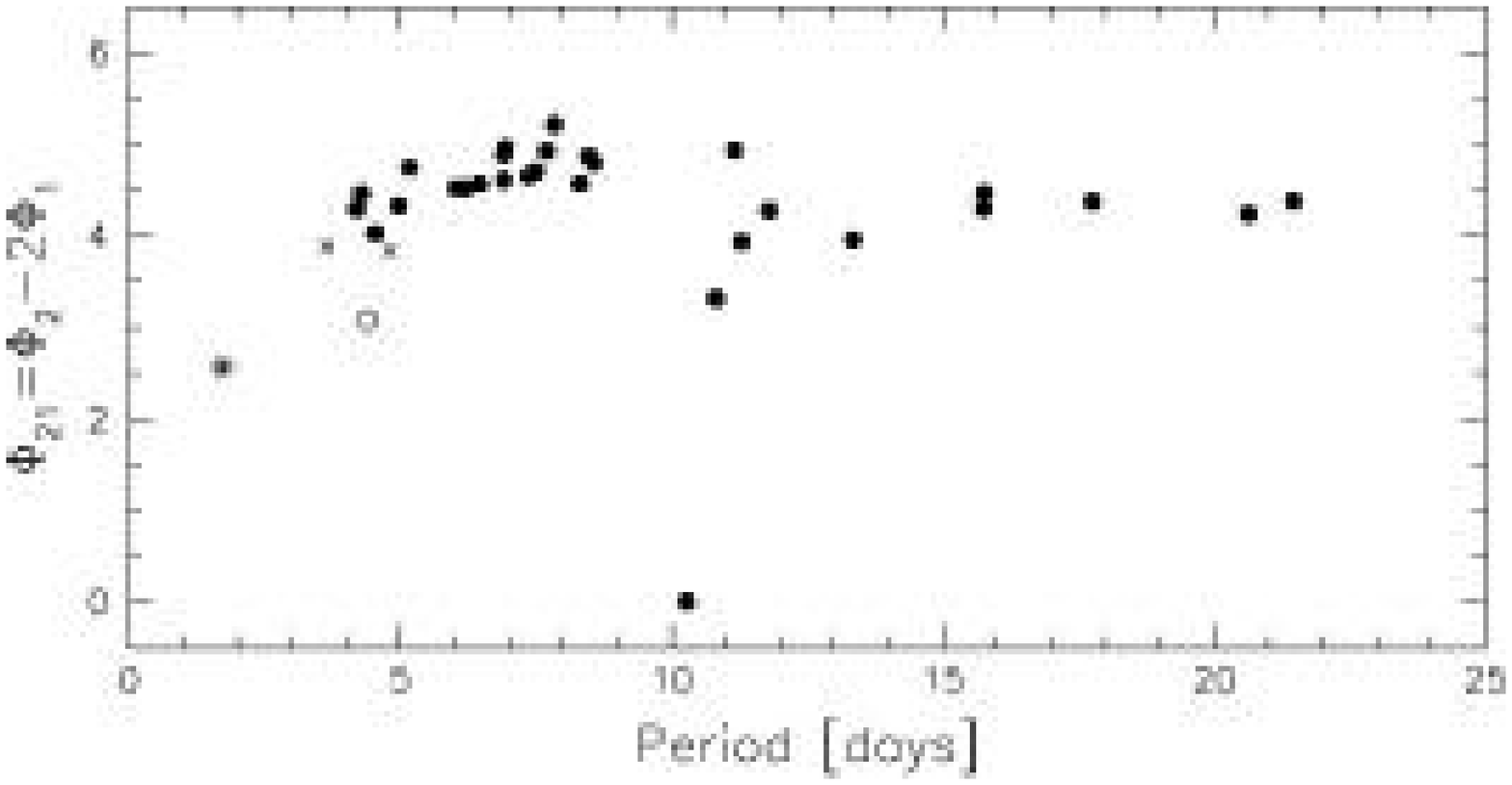}
\caption{\label{fig.fourier1} Amplitude ratio $R_{21}=A_2/A_1$ (upper
panel) and phase difference $\Phi_{21}=\Phi_2 - 2 \Phi_1$ (bottom
panel) determined from the $R$-band data for the population~I Cepheids
(group I) plotted against the period. The ratio and phase difference
drop as the periods approach the resonance $P_2/P_0\approx 0.5$ at
about 10 days and rise again afterwards. The $R_{21}$ value classifies
29 of the detected Cepheids as fundamental mode pulsators (closed
circles), whereas 2 are overtone pulsators (crosses). One of the
Cepheids falls of the fundamental mode sequence in the phase
difference diagram. Following \cite{beaulieu95} we classify this
source as intermediate type Cepheid (open circle). The beat Cepheid
candidate for which we used the dominant period (presumably the first
overtone) in the analysis is marked with an asterisk.}
\end{figure}

We detect 33 population I Cepheids in our sample.  To check whether
our classification is correct and if there are first overtone
pulsators (s-Cepheids) amongst the detected Cepheids we fitted
truncated Fourier series (see Sec.~\ref{sec.fourier}) to the light
curves of group I.  Figure~\ref{fig.fourier1} shows the characteristic
progression of the Fourier parameters with the period of the light
curve for the classical Cepheids: as the amplitude ratio $R_{21}$
between the first and second harmonic drops the corresponding phase
difference shows a mild rise. The amplitude ratio $R_{21}$ declines
until a period of about 10 days is reached, and then starts to rise
again. Due to the suppression of the second harmonic the Cepheid light
curves in the vicinity of this period look quite sinusoidal.
Figure~\ref{fig.fourier1} shows, that the the minimum of $R_{21}$ is
connected to a dramatic change of $\Phi_{21}$. Generally it is
believed that this change is connected to the resonance between the
fundamental mode and the second overtone $P_2/P_0\approx 0.5$ (e.g.,
\citet{simon86}).

First overtone pulsators can be detected using the amplitude ratio
$R_{21}$, as s-Cepheids showing smaller values than their fundamental
mode counterparts.  Following \cite{beaulieu95} we demand $R_{21}<0.3$
for $P < 3$ days and $R_{21}<0.2$ for $3\le P < 5.5$ days for the
Cepheids to be classified as first overtone pulsators. Using this
criterion we identify 2 first overtone pulsators in our sample. One
source with a $R_{21}$ value close to the border to the s-Cepheids
falls of the fundamental mode sequence in the phase difference
diagram. Following \cite{beaulieu95} we classify this source as
intermediate Cepheid.

\begin{figure}[t]
\centering
\includegraphics[width=0.4\textwidth,angle=0]{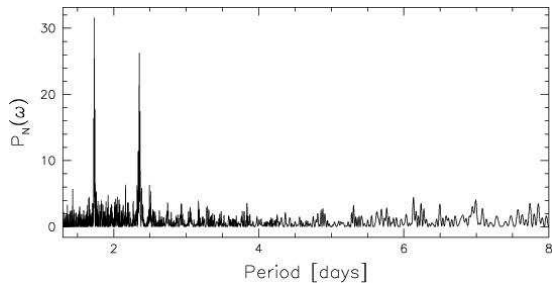}
\caption{\label{fig.beatpower} Power spectrum of the beat
Cepheid candidate in the $R$-band. The two peaks in the spectrum correspond to
periods of 1.73 days and 2.35 days.}
\end{figure}

One source is clearly separated from the Cepheid relation in the
period-amplitude plane (see Fig.~\ref{fig.peramp}). This variable
shows Cepheid like variations which are modulated by a period of 208
days in $R$. The modulation is most likely due to another variable
source inside the PSF.  Fourier decomposition of the $R$-band data
shows that the source has two excited periods, 1.73 days and 2.35
days, besides the subdominant modulation of the light curve.  In
Fig.~\ref{fig.beatpower} we show the power spectrum of this source.
The two periods would classify the source as a beat Cepheid, the first
one detected in M31. Beat Cepheids are a rare sub-class of Cepheids in
which two pulsation modes are excited simultaneously. The ratio of the
two periods $P_{short}/P_{long}\approx 0.73$ makes it likely that we
see a fundamental mode / first overtone (F/1H) pulsator, if the
Cepheid hypothesis for this source is confirmed.  For F/1H beat
Cepheids the first overtone should be the dominant mode, which is
fulfilled for our candidate.  Interestingly, the position in the
$(P_1/P_0)$ - $\log\,(P_0)$ diagram would place this beat Cepheid on a
sequence defined by the SMC beat Cepheids of \cite{beaulieu97}, well
above the relations for the LMC and the Galaxy.  This would point to a
Cepheid in M31 of approximately the same metal content by mass as the
SMC beat Cepheids. The object is resolved in our reference image and
is also classified as a star (261262) in the \cite{haiman}
catalogue. Additionally it was detected in X-rays and appears in the
\citet{kong02} catalogue (J004301.8+411726, see also
Table~\ref{tab.kong}).

In addition to the beat Cepheid there is another peculiar source in
group I which remains unclassified at this stage. It shows two periods
of 6.95 days and 20.5 days. We will further investigate that object,
together with the beat Cepheid candidate, and also present a more
detailed description of the Fourier parameters of group I stars in a
future publication.
\subsubsection{Type II Cepheids and RV Tauri stars}
Type II Cepheids are low-mass population II variables which follow a
period luminosity relation about 1.5 mag below the classical Cepheid
relation.  They are found in old populations like globular
clusters, the halo or the bulge.  Type II Cepheids with periods less
than 5 days are also called BL Her or CVB stars, whereas type II
Cepheids with periods between 10 and 20 days are often referred as W
Virgini stars. At the upper period limit often a period doubling can
be observed in the light curves. An approximate period limit of about
20 days \citep{alcock98} separates type II Cepheids from RV Tauri
stars which share the same light curve, and often also the same
chemical and dynamical characteristics \citep{fokin01}. RV Tauri stars
can be recognized by a typical double-wave light curve with
alternating deep and shallow minima, their semi-periods (minimum to
minimum) ranging from 20 to over 50 days. The current understanding
places RV Tauri stars at the end of stellar evolution. After leaving
the Asymptotic Giant Branch (AGB, \citet{agb}) they move left in the
HR-diagram entering the instability strip at high luminosities. For a
recent review on population II Cepheids and related stars see
\citet{wallerstein02}.  A comprehensive collection of type II Cepheids
and RV Tauri stars detected in the LMC by the MACHO collaboration can
be found in \citet{alcock98}.

\begin{figure}[t]
\centering
\includegraphics[width=0.49\textwidth,angle=0]{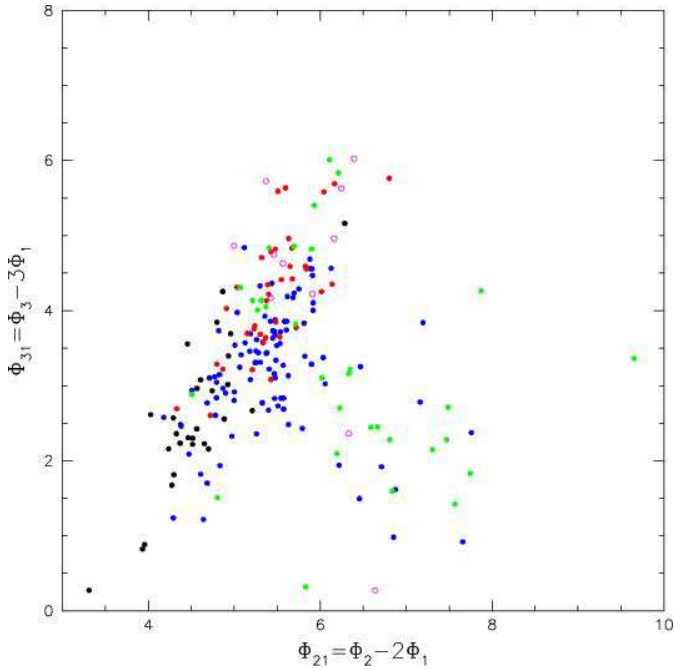}
\caption{{\label{fig.fourier2} Phase differences
$\Phi_{ij}=\Phi_i-i\,\Phi_j$ as determined from the $R$-band data
plotted against each other.  Black dots: population I Cepheids (group
I), blue dots: type II Cepheids (group II), red dots: RV Tauri stars
(group II), open magenta circles: RV Tauri candidates (group II).  As
green dots we show the LMC type II Cepheids and RV Tauri stars from
\citet{alcock98}, which fall on the sequences of the WeCAPP
sources. For the RV Tauri stars we use the formal period in the
analysis. The classical Cepheids show a clear correlation of the two
parameters, but also for the type II Cepheids and the RV Tauri stars
both phase differences are correlated. The sequences of RV Tauri
stars and type II Cepheids overlap, making the RV Tauri sequence to an
extension of the Cepheid II sequence. This favors the close connection
between these two types of stars.  }}
\end{figure}

We detect 37 RV Tauri stars and 11 RV Tauri candidates which makes
this catalogue to one of the biggest collections of RV Tauris to
date. The light curves show the typical alternation of deep and
shallow minima, the second maximum being fainter than the first one,
although for a few sources this latter difference is rather marginal.
The light curve shape can be divided into two groups, one resembling
the {\it `flat-topped'} Cepheid II shape (see
Fig.~\ref{fig.rv_tauri}), the other one showing sinusoidal variations.
To avoid a mis-classification of the sinusoidal light curves of RV
Tauri stars with the not too different light curves of $\beta$-Lyrae
eclipsing binaries, we extract the phase difference between the
second and the fourth harmonic from Fourier fits to the light
curves. According to \cite{szymanski} these two phases should be
strongly coupled for sinusoidal contact systems, yielding $\Delta
\Phi=\Phi_4-2\Phi_2=0$ (M. Szymanski, priv.comm.).  For pulsating
stars the correlation is much weaker. 5 of our RV Tauri candidates
with sinusoidal light curves have a $\Delta \Phi$ which is within
1-$\sigma$ compatible with 0 (or equivalently 2$\pi$), another 6 lie
within 3-$\sigma$. We classify all 11 RV Tauris whose $\Delta \Phi$
values are within 3-$\sigma$ compatible with 0 as RV Tauri candidates,
the remaining 37 sources as RV Tauri stars.

Using the fundamental period (minimum to minimum) for the Fourier
extension of the RV Tauri sources yields relatively bad fits and
subsequently more uncertain Fourier parameters.  We therefore use the
formal period in the analysis which yields a more appropriate
description of the data, better fits, and more reliable Fourier
parameters.  In Fig.~\ref{fig.fourier2} we show a `phase-phase'
diagram as a further result of the Fourier analysis of the RV Tauri
and candidate RV Tauri light curves. The phase differences $\Phi_{31}$
and $\Phi_{21}$ correlate for the RV Tauri light curves (red dots)
showing that these sources form a homogeneous group of stars.  Also
the RV Tauri candidates (magenta open circles) follow the relation
supporting the RV Tauri nature of these stars.  Furthermore we show
the detected type II Cepheids as blue dots in this figure.  The
sequences of RV Tauri stars and type II Cepheids overlap, making the
RV Tauri sequence to an extension of the Cepheid sequence at higher
periods. This supports the close connection between RV Tauri stars and
type II Cepheids. In fact, \cite{alcock98} showed that a single
period-luminosity-color relationship describes both the type II
Cepheids and RV Tauri stars in the LMC. We show the phase parameters
for the stars presented in this study as green dots in Fig. 17.  For
this purpose we re-analyzed the light curves taken from the MACHO project
database\footnote{http://wwwmacho.mcmaster.ca/Data/MachoData.html} and
used the formal period to derive the Fourier parameters. Two stars
with an uncertain classification were rejected from the analysis.  The
LMC variables lie on the sequence of the WeCAPP RV Tauri and Cepheid
II stars, supporting our classification of the variables of
group~II. We will present a more thorough discussion of the Fourier
parameters of the Cepheid-like variables in group I and II in a future
publication.

The period range of the 93 type II Cepheids extends to periods larger
than the approximate limit of about 20 days as proposed by
\citet{alcock98}. There is still the possibility that some of these
long period Cepheids are in reality RV Tauri stars, since noise in the
light curves can prevent us to detect the alternation of deep and
shallow maxima as required for the assignment of the variable as a RV
Tauri candidate or star.  The majority of the light curves of this
type of variables are of `{\it flat-topped}' shape \citep{kwee} with
relatively long and flat maxima. As already mentioned, the light curves
of type II Cepheids show a progression of the phase differences and
occupy distinct places in Fourier space. The sample of type II
Cepheids outnumbers the sample of population I Cepheids by about a
factor of three. As the type II Cepheids trace the old bulge
population this is not unexpected, even when taking into account their
smaller brightness (at the same period) if compared to population~I
Cepheids. The population~I Cepheids, more massive and younger than the
type II Cepheids, are typical members of the M31 disk population,
which is superimposed to the M31 bulge.

The third constituent of the group~II sources are small period
semi-regular variables which contribute from periods about 16 days
onwards. The Fourier analysis of the light curves shows no
correlations of the amplitude ratios or phase differences with
period. Also phase-phase diagrams do not reveal a correlation of the
phase parameters.

\begin{figure}
\centering
\includegraphics[width=0.4\textwidth,angle=0]{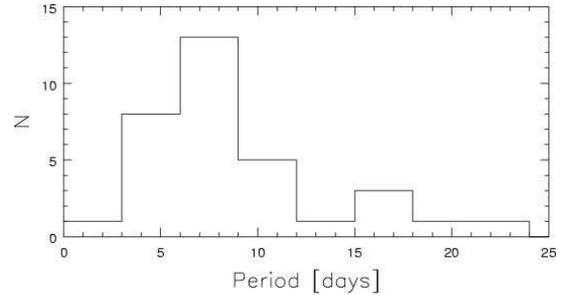}
\caption{\label{fig.phist_g1} Histogram of the periods of the
Cepheid variables in group~I.} 
\end{figure}

\begin{figure}
\centering
\includegraphics[width=0.4\textwidth,angle=0]{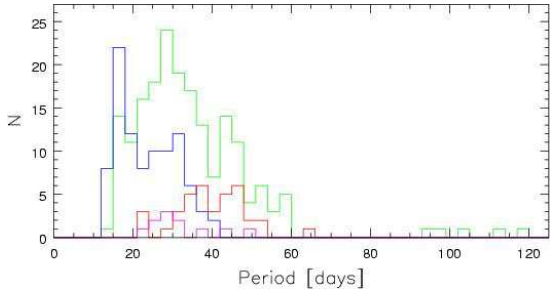}
\caption{\label{fig.phist_g2} Histogram of the periods of variables
belonging to group~II. We show the distributions for type~II Cepheids
(blue line), RV Tauri stars (red line), RV Tauri candidates (magenta
line) and the semi-regular stars (green line). For the RV Tauri stars
and candidates we use the single or fundamental period (minimum to
minimum).}
\end{figure}

In Fig.~\ref{fig.phist_g1} we give the period distribution of the
population~I Cepheids, whereas Fig.~\ref{fig.phist_g2} shows the
distribution of periods for variables belonging to group~II. In red,
magenta, blue, and green we present the distribution for the RV Tauri
stars, RV Tauri candidates, type II Cepheids, and semi-regular
variables.
\subsection{Group III - LPVs}
Group III consists of Long Period Variables (LPVs), i.e.  Mira and
semi-regular variable stars.  LPVs are members of the AGB which marks
the final stage of the stellar evolution for intermediate mass stars
with masses between 0.5 and 8 $M_\odot$.  These stars evolve from the
main sequence and populate the red giant branch up to a maximum
luminosity at the TRGB (tip of the red giant branch). After ignition
of helium core burning they drop in luminosity and form the horizontal
branch. At the end of this stadium the luminosity again is rising as
the stars evolve upwards the AGB.  This evolution sequence is
characterized by pulsation and extensive mass loss.

LPVs are a very promising tool for various astrophysical
questions. They follow tight period-luminosity relations in the
near-IR (especially the K-band) \citep{feast89,wood,feast02} which
makes them excellent galactic and extragalactic distance
estimators. They are furthermore good indicators of the parent
population to which they belong, as there exists a dependence of the
luminosity (hence period) of Miras on the age. Longer period Miras
should have higher mass progenitors and therefore belong to a younger
population.  LPVs are historically separated in two main groups: i)
Miras with regular variations, periods between 80 and 1000 days and an
amplitude of the variation in the $V$-band of more than 2.5 mag; ii)
semi-regulars (SR) with less regular variations, smaller periods and a
$V$-band variation smaller than 2.5 mag. Semi-regulars are dived in
two groups, SRas with more regular variations and SRbs with variations
less regular.

\begin{figure}[]
\centering
\includegraphics[width=0.4\textwidth,angle=0]{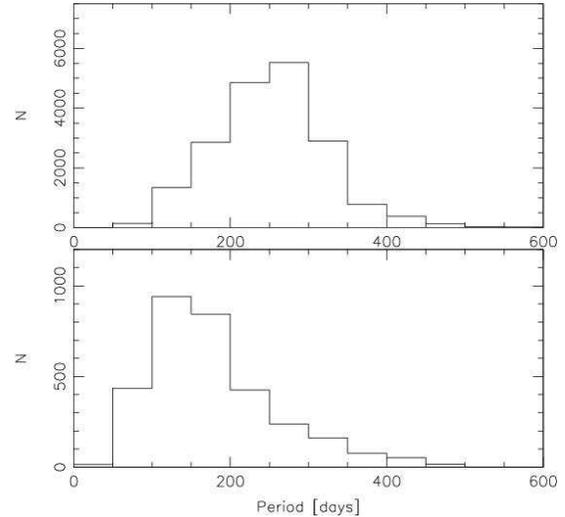}
\caption{\label{fig.pdis} Distribution of the periods of LPVs in the
bulge of M31 detected by the WeCAPP project. Top: class i - regular
and semi-regular variables (significance level $P(>z)$ $<$ $10^{-20}$
in $R$ or $I$).  Bottom: class ii - irregular variables (significance
level $P(>z)$ $\ge$ $10^{-20}$ in $R$ and $I$). Only irregulars
selected by the significance cut are shown.  Note that the
distribution for the sources classified as irregular variables is of
qualitative nature only as the periods for these sources are not well
determined. In general irregular variables show smaller `periods' than
the group of regular and semi-regular variables.  }
\end{figure}

Recently \citet{lebzelter02} proposed a new classification scheme less
dependent on this kind of artificial division and used it for the
AGAPEROS survey of variable red stars towards the Magellanic
Clouds. Their classification is based on the regularity of the
variation alone, providing three classes of stars: i) LPVs with
regular variation ii) LPVs with semi-regular variation iii) LPVs with
irregular variation.  No cuts in amplitude were applied, therefore
class i contains members of the classical Mira group as well as
members of the SRa type of objects.

We modify this classification scheme for our purposes and define a
significance cut of $10^{-20}$ according to the period finding
algorithm in $R$ or $I$ for the division of our sample of LPVs in
classes i (regular and semi-regular) and ii (irregular). Class ii
coincides with the class of irregular stars introduced in
Sec.~\ref{sec.variables}. Because of the non-linear dependence of the
significance on the $S/N$ of the light curve (see also discussion
below) we abandon to invent an automatic cut for the division into
semi-regular and regular variations. The cut at $10^{-20}$ is somewhat
subjective as also stars with semi-regular, but low $S/N$ light curves
are classified as irregular, as the low $S/N$ prevents the period to
be determined with better significance. On the other hand this cut
ensures that most of the irregular light curves are classified in the
right way. The histograms of the periods for the irregular and the
regular/semi-regular sample of LPVs are shown in
Fig. \ref{fig.pdis}. The irregular variables show in the mean smaller
`periods' as the group of regular or semi-regular variables. Note,
that this statement is of qualitative nature only, as the periods for
these sources are not well determined.

\begin{figure}[t]
\centering
\includegraphics[width=0.4\textwidth,angle=0]{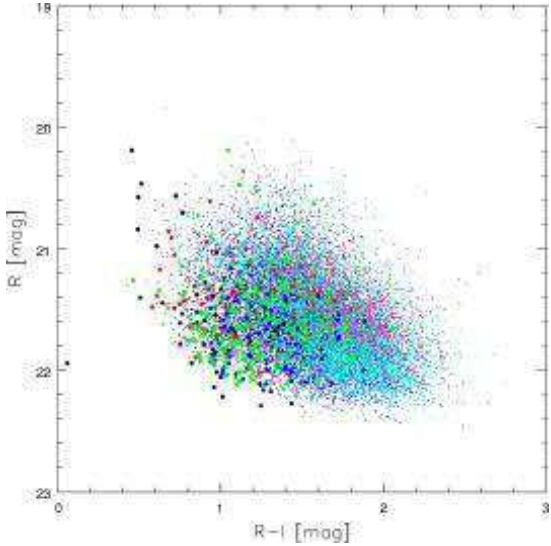}
\caption{\label{fig.colmag} Variation color $R-I$ shown as a function
of the $R$-band variation magnitude. The sources of groups I and II
are shown as big closed circles, the sources of group III as small
closed circles. The color coding is the same as in
Fig.~\ref{fig.peramp}, i.e. the LPVs are again color coded according
to the significance of the $I$-band period. The sources in group I and
II as a whole show bluer variations than the LPVs in group~III. The
significance of the $I$-band period rises for brighter $R$ variation
magnitudes and redder variation colors. This is a result of the higher
$S/N$ of the $I$-band light curves, but can in part also be attributed
to the enhanced regularity of the light curves.  }
\end{figure}
 
\begin{figure}[]
\centering
\includegraphics[width=0.4\textwidth,angle=0]{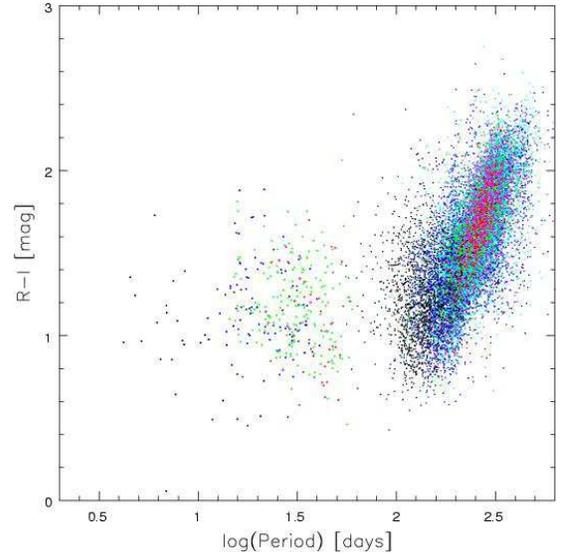}
\caption{\label{fig.percol} This figure shows the variation color
$R-I$ as function of the period of the variable sources. The color
coding is the same as in Figs.~\ref{fig.peramp} and \ref{fig.colmag}.
Note, that due to the normalization to the fluxes of Vega in the
particular system this $R-I$ color has not to be mistaken with the
color of the variation amplitudes of the sources. Nevertheless, some
trends are visible in this diagram: As the variation of the LPVs gets
redder with increasing period, the $I$-band periods become more
significant, reflecting the rising $S/N$ in $I$ due to higher $R-I$
values and a rising regularity of the light curves.}
\end{figure}

We have color coded the LPVs in Fig.~\ref{fig.peramp} according to the
significance of the $I$-band period. The slope of the variation
magnitude-period relation changes from slightly positive for the low
significance tail (black dots) to negative for the sources with
intermediate and high significant periods (green and red dots).  The
same trend can be seen in Fig.~\ref{fig.colmag}, which shows the
variation color $R-I$ as a function of the $R$-band variation
magnitude. The significance of the $I$-band period rises for brighter
$R$ variation magnitudes and redder variation colors. The differences
in the significance of the LPV $I$-band periods therefore is a result
of the different $S/N$ of the $I$-band light curves (rising $S/N$ in
$I$ due to higher $R-I$ values), but can in part be also attributed to
different stages of regularity in the light
curves. Figure~\ref{fig.percol} finally shows the variation color
$R-I$ as function of the period of the variable sources. The variation
of the LPVs gets redder with increasing period, and at the same time
shows more significant periods.

\section{The Catalogue\label{sec.catalogue}}
The final catalogue comprises 23781 entries with group~I containing
one beat Cepheid candidate, 2~s-Cepheids, one Cepheid of intermediate
type, and 29 classical Cepheids pulsating in the fundamental mode. One
further group I source which remains unclassified in this catalogue
shows two periods.

Group~II comprises 93 type~II Cepheids, 37 RV Tauri stars, 11 RV Tauri
candidates, and 193 low period semi-regular variables.  Group~III
consists of 4287 irregular and 18974 regular/semi-regular
variables. 82 presumably group~III members show variations on
timescales longer than the survey length. The positions in the
$R$-band period-amplitude plane of the 31 eclipsing binary candidates
coincide with the positions of groups I and II. Finally we detect 39
miscellaneous variables, among them 16 novae and 15 R Coronae Borealis
candidates.

\begin{figure}[t]
\begin{center}
\includegraphics[width=0.4\textwidth,angle=0]{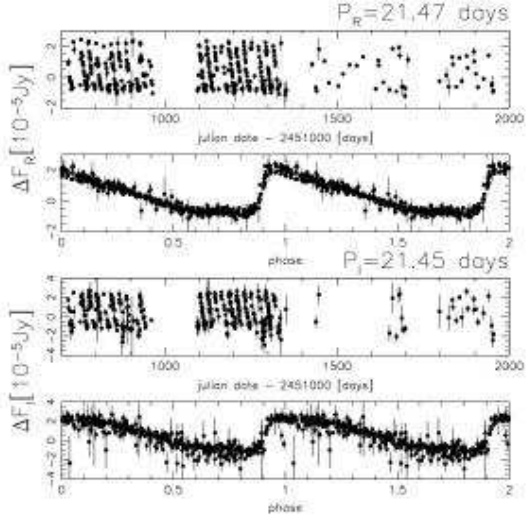}
\end{center}
\caption{\label{fig.d_ceph} Light curve of a $\delta$-Cepheid
(group~I) in the $R$-band (top panels) and the $I$-band (bottom
panels). In the upper panels we show the light curve $R$-band, the
lower panels show the light curves in the $I$-band. In both bands the
light curves are presented in the time domain (top for each band) and
and in the phase domain, i.e. convolved with the derived periods
(bottom for each band). Additionally we show the periods $P_R$ and
$P_I$ derived in the $R$- and $I$-band, respectively.}
\end{figure}

\begin{figure}[]
\begin{center}
\includegraphics[width=0.4\textwidth,angle=0]{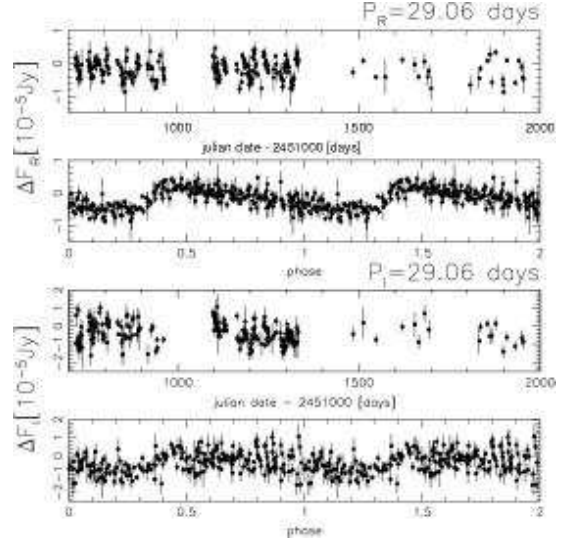}
\end{center}
\caption{\label{fig.II_ceph} Light curve of a long period type~II
Cepheid (group~II) in the $R$-band (top panels) and the $I$-band
(bottom panels).}
\end{figure}

\begin{figure}[]
\begin{center}
\includegraphics[width=0.4\textwidth,angle=0]{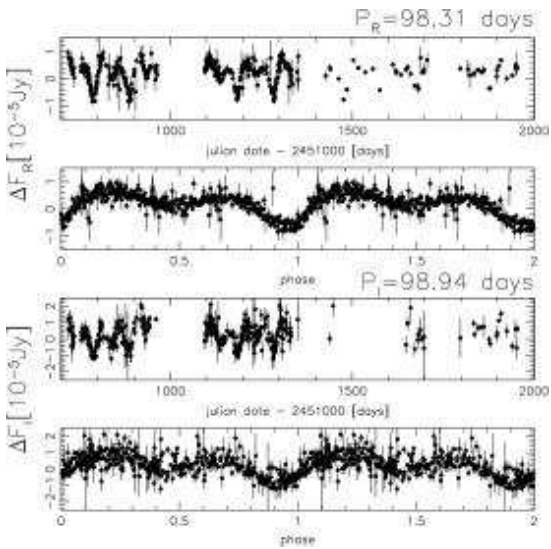}
\end{center}
\caption{\label{fig.rv_tauri} Light curve of a RV Tauri star
(group~II) in the $R$-band (top panels) and the $I$-band (bottom
panels). We show the light curves convolved with the formal periods
(deep minimum to deep minimum). The double wave light curve with
alternating deep and shallow minima is nicely uncovered.}
\end{figure}

\begin{figure}[]
\begin{center}
\includegraphics[width=0.4\textwidth,angle=0]{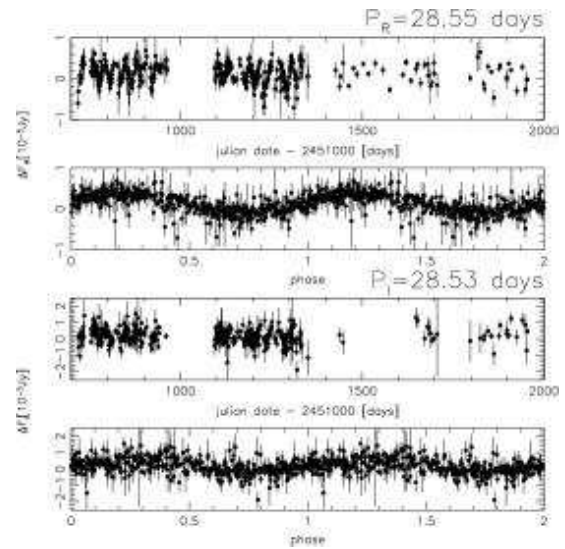}
\end{center}
\caption{\label{fig.sr_g2} Light curve of a low period semi-regular
variable (group~II) in the $R$-band (top panels) and the $I$-band
(bottom panels).}
\end{figure}

\begin{figure}[]
\begin{center}
\includegraphics[width=0.4\textwidth,angle=0]{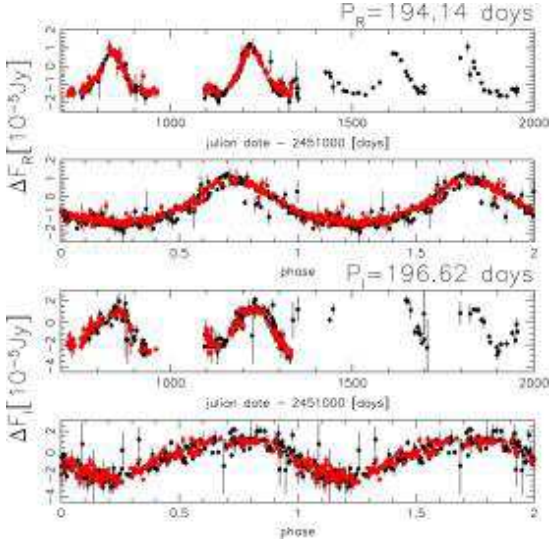}
\end{center}
\caption{\label{fig.reg_lpv} Light curve of a LPV with regular
variation (group~III) in the $R$-band (top panels) and the $I$-band
(bottom panels).  As illustration of the agreement of the data taken
with different telescopes we show the data collected at Calar Alto as
red dots, and data taken at Wendelstein as black dots.}
\end{figure}

\begin{figure}[]
\begin{center}
\includegraphics[width=0.4\textwidth,angle=0]{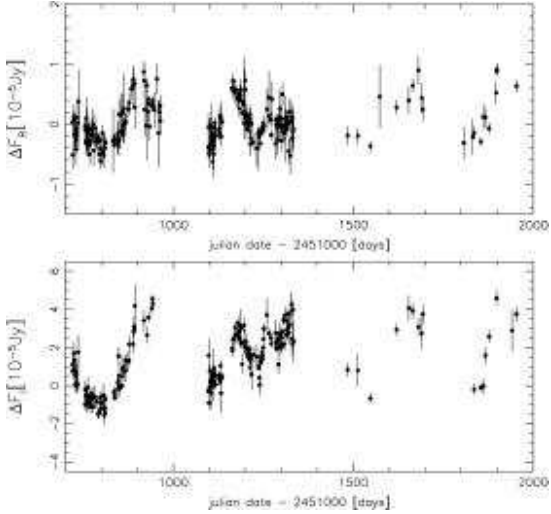}
\end{center}
\caption{\label{fig.irr_lpv} Light curve of an irregular LPV
(group~III).  Top panel: $R$-band. Bottom panel: $I$-band.}
\end{figure}

\begin{figure}[]
\begin{center}
\includegraphics[width=0.4\textwidth,angle=0]{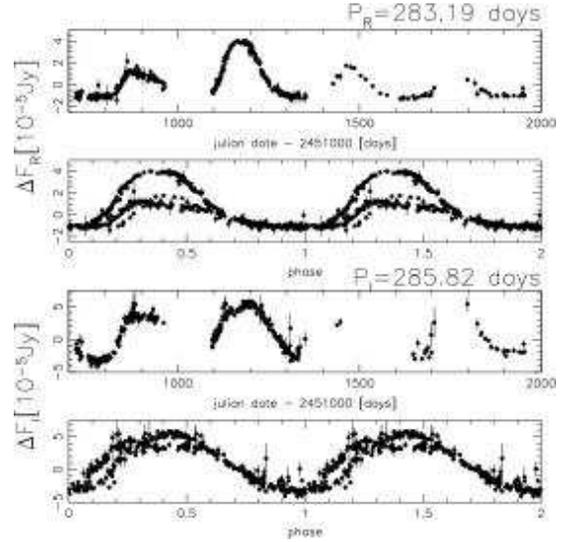}
\end{center}
\caption{\label{fig.amp_lpv} Light curve of a LPV (group~III) with
changing variation amplitudes from cycle to cycle. Top panels:
$R$-band.  Bottom panels: $I$-band.}
\end{figure}

\begin{figure}[]
\begin{center}
\includegraphics[width=0.4\textwidth,angle=0]{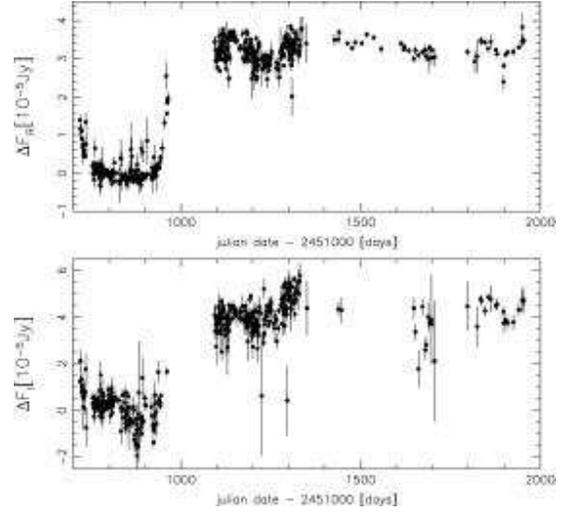}
\end{center}
\caption{\label{fig.rcb} Light curve of a R Coronae Borealis
candidate (miscellaneous variables). Top panel: $R$-band. Bottom
panel: $I$-band. These rare carbon-rich stars show unpredictable and
frequent fading of light in optical wavebands, most probably due to
dust grains ejected from the stellar photosphere. For a review on
these RCB stars see \citet{clayton96}.}
\end{figure}

\begin{figure}[]
\begin{center}
\includegraphics[width=0.4\textwidth,angle=0]{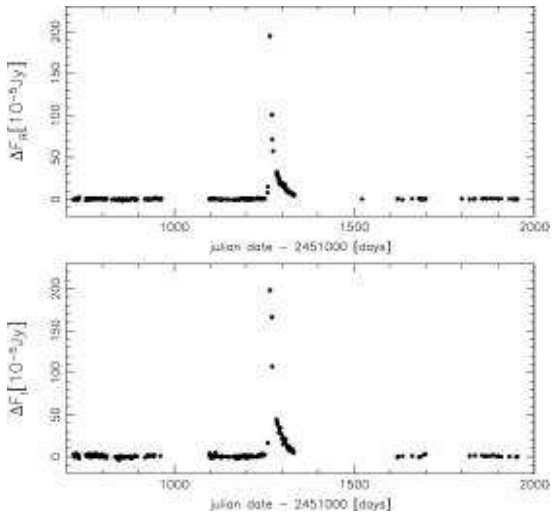}
\end{center}
\caption{\label{fig.nova} Light curve of a nova (miscellaneous
variables). Top panel: $R$-band. Bottom panel: $I$-band.  A paper
concerning the characteristics of all detected novae is in
preparation.}
\end{figure}

The full catalog is available in electronic form at CDS.  We give the
name, the position in the WCS, the periods (if available) derived from
the $R$- and $I$-band data, the amplitudes of the variation $\Delta
F_R$ and $\Delta F_I$, and provide a classification according to
Secs. \ref{sec.variables} and \ref{sec.classes}: DC (classical or
$\delta$-Cepheids), SC (s-Cepheids), BC (beat Cepheids), IC
(intermediate Cepheids), W (type II Cepheids), RV (RV Tauri stars), S
(regular or and semi-regular variations), and I (irregular
variables). We also mark the identified Novae (N), eclipsing binary
candidates (E), RCB candidates (RCB) and other miscellaneous variables
(M). As an illustration of its content we list in
Table~\ref{tab.catalogue} the entries 100 $-$ 120 in the catalogue. In
Figs.~\ref{fig.d_ceph} to \ref{fig.nova} we show typical light curves
for each of the groups.

\begin{table*}[t]
\begin{center}
\begin{tabular}{crrccccccc}
\hline \hline
Id & $\alpha(2000.0)$ & $\delta(2000.0)$ & $P_R$[days] & -$\log(s_R)$ & $P_I$[days]  & -$\log(s_I)$ & 
$\Delta F_R$ & $\Delta F_I$ & Type\\
\hline
WeCAPP\_V00100  &  0h42m18.46s  &  41d08$\arcmin$09.0$\arcsec$  &  119.70  &  13.18  &  122.10  &   8.52  &  0.37  &  1.06  &  I  \\
WeCAPP\_V00101  &  0h42m09.65s  &  41d08$\arcmin$08.1$\arcsec$  &  274.99  &  21.65  &  277.50  &  26.10  &  0.78  &  2.86  &  S  \\
WeCAPP\_V00102  &  0h42m31.76s  &  41d08$\arcmin$11.5$\arcsec$  &  500.40  &   9.60  &  478.70  &  22.69  &  0.47  &  3.48  &  S  \\
WeCAPP\_V00103  &  0h42m20.06s  &  41d08$\arcmin$04.4$\arcsec$  &  180.79  &  29.32  &  180.21  &  26.71  &  0.76  &  1.70  &  S  \\
WeCAPP\_V00104  &  0h42m31.53s  &  41d08$\arcmin$08.1$\arcsec$  &  289.96  &  18.52  &  293.51  &  23.71  &  0.74  &  3.18  &  S  \\
WeCAPP\_V00105  &  0h42m52.12s  &  41d08$\arcmin$11.0$\arcsec$  &  299.24  &  33.60  &  302.18  &  41.93  &  0.63  &  2.59  &  S  \\
WeCAPP\_V00106  &  0h43m17.73s  &  41d08$\arcmin$12.8$\arcsec$  &   136.3  &  8.920  &  140.74  &   4.86  &  0.49  &  1.05  &  I  \\
WeCAPP\_V00107  &  0h42m19.93s  &  41d08$\arcmin$09.7$\arcsec$  &  284.59  &  10.99  &  273.54  &  29.43  &  0.43  &  2.03  &  S  \\
WeCAPP\_V00108  &  0h42m45.51s  &  41d08$\arcmin$10.7$\arcsec$  &  268.01  &  19.64  &  267.94  &  38.87  &  0.57  &  2.56  &  S  \\
WeCAPP\_V00109  &  0h42m18.84s  &  41d08$\arcmin$08.8$\arcsec$  &  240.35  &  21.15  &  247.07  &  27.50  &  0.51  &  1.74  &  S  \\
WeCAPP\_V00110  &  0h42m17.75s  &  41d08$\arcmin$08.8$\arcsec$  &  423.96  &  5.730  &  423.96  &  24.32  &  0.48  &  1.36  &  S  \\
WeCAPP\_V00111  &  0h42m17.43s  &  41d08$\arcmin$08.6$\arcsec$  &  280.04  &  16.27  &  278.51  &  27.95  &  0.48  &  1.69  &  S  \\
WeCAPP\_V00112  &  0h42m30.69s  &  41d08$\arcmin$07.9$\arcsec$  &  216.48  &  17.07  &  218.03  &  21.92  &  0.81  &  2.21  &  S  \\
WeCAPP\_V00113  &  0h42m17.04s  &  41d08$\arcmin$06.8$\arcsec$  &  229.50  &  25.36  &  229.51  &  28.31  &  0.79  &  2.58  &  S  \\
WeCAPP\_V00114  &  0h43m02.66s  &  41d08$\arcmin$13.0$\arcsec$  &  226.44  &  29.33  &  230.99  &  35.27  &  0.53  &  1.96  &  S  \\
WeCAPP\_V00115  &  0h42m35.32s  &  41d08$\arcmin$12.0$\arcsec$  &     -    &    -    &     -    &     -   &     -  &    -   &  L  \\
WeCAPP\_V00116  &  0h42m23.24s  &  41d08$\arcmin$11.1$\arcsec$  &  314.68  &  24.78  &  319.13  &  31.99  &  0.53  &  2.60  &  S  \\
WeCAPP\_V00117  &  0h42m14.70s  &  41d08$\arcmin$10.1$\arcsec$  &  214.96  &  26.14  &  209.84  &  27.26  &  0.59  &  1.77  &  S  \\
WeCAPP\_V00118  &  0h42m24.75s  &  41d08$\arcmin$12.4$\arcsec$  &  277.49  &  10.20  &  280.04  &  23.22  &  0.42  &  1.53  &  S  \\
WeCAPP\_V00119  &  0h42m47.43s  &  41d08$\arcmin$11.0$\arcsec$  &  497.12  &  24.86  &  505.28  &  28.47  &  0.80  &  6.19  &  S  \\
WeCAPP\_V00120  &  0h42m24.29s  &  41d08$\arcmin$11.5$\arcsec$  &  198.21  &  21.46  &  198.61  &  23.74  &  0.46  &  2.82  &  S  \\
\hline
\end{tabular}
\caption{\label{tab.catalogue} Extract from the WeCAPP catalogue of
variable stars.  We give the identification, the right ascension and
declination, the assigned period and its significance $s$ ($0.1
\stackrel{\wedge}{=}$ un-significant, $70 \stackrel{\wedge}{=}$ highly
significant) in $R$ and $I$, the amplitude of variation in both bands,
and finally a classification according to the scheme presented in
Secs.~\ref{sec.variables} and \ref{sec.classes}. The variation amplitudes
$\Delta F_R$ and $\Delta F_I$ are given in units of [10$^{-5}$ Jy].
}
\end{center}
\end{table*}

\section{Correlation with other catalogues\label{sec.comparison}}
We cross-correlate our $R-$band selected catalogue with the General
Catalogue of Variable Stars (GCVS, \cite{durlevich}) and catalogues
selected in the $X$-ray (Chandra; \citet{kaaret02} \& \citet{kong02}).

We find 23 coincidences with the \citet{kaaret02} catalogue when using
a search radius of 1$\arcsec$. To estimate how many coincidences we
expect by chance we follow the approach proposed by
\citet{hornschemeier}.  We shift one of the catalogues to be compared
(in our case we shift the catalogues taken from the literature) by
20$\arcsec$ in north-east, south-east, south-west and north-west
direction, and check for coincidences between these mock catalogues
and our catalogue. By averaging these numbers we get the expected
number of coincidences by chance.

\begin{table*}
\centering
\begin{tabular}{rrrccc}
\hline\hline
WeCAPP &  $\alpha$(2000) & $\delta$(2000) & Kaaret & 2MASS & $\Delta$ r(arcsec) \\
\hline
 10578  &  0h42m10.30s  &   41d15$\arcmin$10.4$\arcsec$  &   J004210.2+411510       & 0042102+411510   &  0.5      \\
 16262  &  0h42m12.14s  &   41d17$\arcmin$58.8$\arcsec$  &   J004212.1+411758       & 0042121+411758   &  0.4      \\
  8624  &  0h42m18.68s  &   41d14$\arcmin$01.8$\arcsec$  &   J004218.6+411402       & 0042186+411402   &  0.5      \\
  8946  &  0h42m21.58s  &   41d14$\arcmin$19.8$\arcsec$  &   J004221.5+411419       & 0042215+411419   &  0.4      \\
  7582  &  0h42m25.11s  &   41d13$\arcmin$40.7$\arcsec$  &   J004225.1+411340       &                  &  0.6      \\
 19068  &  0h42m31.26s  &   41d19$\arcmin$38.7$\arcsec$  &   J004231.2+411938       & 0042312+411938   &  0.2      \\
  7979  &  0h42m36.59s  &   41d13$\arcmin$49.9$\arcsec$  &   J004236.6+411350       &           &  0.7      \\
  9082  &  0h42m39.60s  &   41d14$\arcmin$28.6$\arcsec$  &   J004239.5+411428       &           &  0.9      \\
 11406  &  0h42m41.45s  &   41d15$\arcmin$23.9$\arcsec$  &   J004241.4+411524       &           &  0.3      \\
 16322  &  0h42m43.97s  &   41d17$\arcmin$55.5$\arcsec$  &   J004243.9+411755       &           &  0.4      \\
 15344  &  0h42m44.88s  &   41d17$\arcmin$39.5$\arcsec$  &   J004244.8+411740       &           &  0.4      \\
 12503  &  0h42m46.99s  &   41d16$\arcmin$15.3$\arcsec$  &   J004246.9+411615       &           &  0.7      \\
  8501  &  0h42m47.13s  &   41d14$\arcmin$13.9$\arcsec$  &   J004247.1+411413       &           &  0.8      \\
 10431  &  0h42m47.45s  &   41d15$\arcmin$07.6$\arcsec$  &   J004247.4+411507       &           &  0.2      \\
 11736  &  0h42m47.90s  &   41d15$\arcmin$50.6$\arcsec$  &   J004247.8+411550       &           &  0.6      \\
 12891  &  0h42m48.65s  &   41d16$\arcmin$25.0$\arcsec$  &   J004248.6+411624       &           &  0.7     \\
 20170  &  0h42m55.27s  &   41d20$\arcmin$45.1$\arcsec$  &   J004255.3+412045       &           &  0.7     \\
 18374  &  0h42m59.66s  &   41d19$\arcmin$19.3$\arcsec$  &   J004259.6+411919       &           &  0.3      \\
 12404  &  0h42m59.88s  &   41d16$\arcmin$06.0$\arcsec$  &   J004259.8+411606       &           &  0.1      \\
 21059  &  0h43m03.31s  &   41d21$\arcmin$21.8$\arcsec$  &   J004303.2+412121       &           &  0.5      \\
  6049  &  0h43m08.42s  &   41d12$\arcmin$46.9$\arcsec$  &   J004308.4+411247       &           &  0.7      \\
  8360  &  0h43m09.88s  &   41d19$\arcmin$00.8$\arcsec$  &   J004309.8+411900       &           &  0.7     \\
  9991  &  0h43m10.62s  &   41d14$\arcmin$51.3$\arcsec$  &   J004310.5+411451       & 0043106+411451 &   0.5     \\
\hline
\end{tabular}
\caption{\label{tab.kaaret} Coincidences between the WeCAPP catalogue
of variable stars and the X-ray selected catalogue of point sources by
\citet{kaaret02}. A search radius of 1$\arcsec$ was used. We give the
identifier and WCS coordinates for the WeCAPP sources, and the
identifier for the \citet{kaaret02} sources. For sources coincident
with the 2MASS catalogue \citep{cutri00} we give the 2MASS identifier
also. Additionally we show for each of the correlated sources the
difference $\Delta$ r of the matching of both catalogues. Source
WeCAPP\_V10431 is identical with the (dwarf-)nova WeCAPP-N2000-05 which
shows supersoft X-ray emission \citep{pietsch05}.}
\end{table*}

Applying a search radius of 1$\arcsec$ the number of false
coincidences with the \citet{kaaret02} catalogue becomes 12.  This
high level of false coincidences (about 50\%) results from the high
number density of the variable star catalogue.  Nevertheless, the
false detections rate suggests that about 11 of the coincidences
should be real. This is supported by the fact, that 6 of the
coincident sources have an entry in the 2 MASS catalogue
\citep{cutri00}, 8 of the Kaaret-WeCAPP sources are identified as
globular clusters (4 of them with a 2MASS entry, \citet{kaaret02}).
Furthermore we find that 12 of the sources have coincident
counterparts in the \citet{kong02} catalogue as well (see below).  In
Table~\ref{tab.kaaret} we give the positions of all 23 coincidences,
for sources with entries in the 2MASS catalogue we give the 2MASS name
also. Two of the Kaaret-WeCAPP sources with 2MASS fluxes (WeCAPP\_V8946
and WeCAPP\_V10578) have no globular cluster (gc) counterpart
\citep{kaaret02}. Whereas V8946 coincides with a gc-candidate
identified by \citet{wirth} a gc-counterpart for WeCAPP\_V10578 is
completely unknown.

\begin{table*}[]
\centering
\begin{tabular}{rrrcr}
\hline\hline
WeCAPP & $\alpha$(2000) & $\delta$(2000) & Kong et al.   & $\Delta$ r(arcsec) \\
\hline
14906  &  0h42m07.08s  &   41d17$\arcmin$20.2$\arcsec$  &  J004207.0+411719    &     0.9  \\
15982  &  0h42m09.49s  &   41d17$\arcmin$45.6$\arcsec$  &  J004209.4+411745    &     0.7  \\
16262  &  0h42m12.14s  &   41d17$\arcmin$58.8$\arcsec$  &  J004212.1+411758    &     0.8  \\
 5801  &  0h42m15.09s  &   41d12$\arcmin$34.8$\arcsec$  &  J004215.0+411234    &     0.6  \\
14846  &  0h42m15.65s  &   41d17$\arcmin$21.9$\arcsec$  &  J004215.6+411721    &     0.7  \\
12198  &  0h42m16.06s  &   41d15$\arcmin$53.3$\arcsec$  &  J004216.0+411552    &     0.4  \\
 7582  &  0h42m25.11s  &   41d13$\arcmin$40.7$\arcsec$  &  J004225.1+411340    &     0.3  \\
20282  &  0h42m27.61s  &   41d20$\arcmin$49.0$\arcsec$  &  J004227.6+412048    &     0.3  \\
13905  &  0h42m30.23s  &   41d16$\arcmin$54.3$\arcsec$  &  J004230.2+411653    &     0.9  \\
19068  &  0h42m31.26s  &   41d19$\arcmin$38.7$\arcsec$  &  J004231.2+411939    &     0.7  \\
11863  &  0h42m32.51s  &   41d15$\arcmin$45.7$\arcsec$  &  J004232.5+411545    &     0.2  \\
10895  &  0h42m34.76s  &   41d15$\arcmin$22.9$\arcsec$  &  J004234.7+411523    &     0.3  \\
 7979  &  0h42m36.59s  &   41d13$\arcmin$49.9$\arcsec$  &  J004236.5+411350    &     0.3  \\
11406  &  0h42m41.45s  &   41d15$\arcmin$23.9$\arcsec$  &  J004241.4+411523    &     0.3  \\
10626  &  0h42m43.74s  &   41d15$\arcmin$14.4$\arcsec$  &  J004243.7+411514    &     0.8  \\
15344  &  0h42m44.88s  &   41d17$\arcmin$39.5$\arcsec$  &  J004244.8+411739    &     0.3  \\
15318  &  0h42m46.12s  &   41d17$\arcmin$36.3$\arcsec$  &  J004246.0+411736    &     0.8  \\
12503  &  0h42m46.99s  &   41d16$\arcmin$15.3$\arcsec$  &  J004246.9+411615    &     0.6  \\
11736  &  0h42m47.90s  &   41d15$\arcmin$50.6$\arcsec$  &  J004247.8+411549    &     0.8  \\
22406  &  0h42m48.94s  &   41d24$\arcmin$06.1$\arcsec$  &  J004248.9+412406    &     1.0  \\
17309  &  0h42m56.93s  &   41d18$\arcmin$44.4$\arcsec$  &  J004256.9+411844    &     0.2  \\
18374  &  0h42m59.66s  &   41d19$\arcmin$19.3$\arcsec$  &  J004259.6+411919    &     0.4  \\
12404  &  0h42m59.88s  &   41d16$\arcmin$06.0$\arcsec$  &  J004259.8+411606    &     0.2  \\
15035  &  0h43m01.80s  &   41d17$\arcmin$26.5$\arcsec$  &  J004301.8+411726    &     0.5  \\
10871  &  0h43m02.99s  &   41d15$\arcmin$22.3$\arcsec$  &  J004302.9+411522    &     0.6  \\
21059  &  0h43m03.31s  &   41d21$\arcmin$21.8$\arcsec$  &  J004303.3+412122    &     0.6  \\
16325  &  0h43m03.90s  &   41d18$\arcmin$04.7$\arcsec$  &  J004303.8+411805    &     0.5  \\
19890  &  0h43m07.55s  &   41d20$\arcmin$20.0$\arcsec$  &  J004307.5+412020    &     0.1  \\
17815  &  0h43m09.88s  &   41d19$\arcmin$00.8$\arcsec$  &  J004309.8+411901    &     0.4  \\
 9991  &  0h43m10.62s  &   41d14$\arcmin$51.3$\arcsec$  &  J004310.6+411451    &     0.8  \\
17140  &  0h43m16.14s  &   41d18$\arcmin$41.5$\arcsec$  &  J004316.1+411841    &     0.6  \\
\hline
\end{tabular}
\caption{\label{tab.kong} Coincidences between the WeCAPP catalogue of
variable stars and the X-ray selected catalogue of point sources by
\citet{kong02}. A search radius of 1$\arcsec$ was used. We give the
identifier and WCS coordinates for the WeCAPP sources, and the
identifier for the \citet{kong02} sources. Additionally we show for
each of the correlated sources the difference $\Delta$ r of the
matching of both catalogues. The position of the beat Cepheid
candidate WeCAPP\_15035 coincides with the X-ray source
J004301.8+411726.}
\end{table*}

The \citet{kong02} catalogue contains 31 coincidences with our
variable star catalogue when applying a 1$\arcsec$ search radius. As
we expect 17 coincidences by chance about 14 of the coincidences
should be real. As mentioned above 12 of the sources are also
coincident with the \citet{kaaret02} catalogue. We show the position
of all 31 coincidences in Table~\ref{tab.kong}.
 
The GCVS \citep{durlevich} comprises 250 entries within our field of
view, most of them being of nova type. 161 (6) were classified as
novae, additionally 53 as novae of fast type (NA), 9 (2) of slow type
(NB), and (2) of very slow type (NC). Furthermore the GCVS reports 1
SNIa remnant in our field, 1 (1) LC type (irregular variable
supergiant), 3 (1) semi-regular variables (SR), 3 $\delta$-Cepheids,
(1) SDOR variable, 2 (1+2) irregular variables, and 1 unstudied (S:
classification) variable.  The numbers in brackets give additional
candidates for the particular variable star group with questionable
classifications according to the GCVS.

\begin{table*}[]
\centering
\begin{tabular}{rrrcrrrrr}
\hline\hline
WeCAPP & $\alpha$(2000) & $\delta$(2000) & GCVS   & $\Delta$ r(arcsec) 
& GCVS class& WeCAPP class\\
\hline
906  &	  0h43m22.50s  &   41d07$\arcmin$28.6$\arcsec$  &    V0871   &     0.7 &I:   & I \\
2375 &	  0h43m15.73s  &   41d07$\arcmin$54.7$\arcsec$  &    V0727   &     0.1 &IA:  & I  \\
559  &	  0h43m24.16s  &   41d07$\arcmin$12.7$\arcsec$  &    V0743   &     0.2 &DCEP & DC \\
742  &	  0h43m17.27s  &   44d15$\arcmin$58.0$\arcsec$  &    V0726   &     1.0 &SR   & I \\
\hline
\end{tabular}
\caption{\label{tab.gcvs} Coincidences between the WeCAPP catalogue of
variable stars and the General Catalogue of Variable Stars (GCVS,
\cite{durlevich}). A search radius of 1$\arcsec$ was used. We give the
identifier, WCS coordinates and a classification for the WeCAPP
sources, and the identifier and a classification according to the
GCVS. Additionally we show for each of the correlated sources the
difference $\Delta$ r of the matching of both catalogues.}
\end{table*}

The crowding of the WeCAPP sources demands that we keep the search
radius at 1$\arcsec$, even when comparing with the GCVS, whose
accuracy is well below this value. We find 27 coincidences, the number
of false detections (21) suggests however, that most of the
coincidences are not real. As furthermore 23 of them are classified as
novae in the GCVS we only show the 4 coincidences which we regard as
real in Table~\ref{tab.gcvs}. Note, that we do not match the positions
of all three $\delta$-Cepheids found in the GCVS inside our field. The
GCVS Cepheid V0934 has a position difference of 2.1$\arcsec$ when
compared to our position, GCVS Cepheid V0811 is off by
10.2$\arcsec$. Because of these in parts large uncertainties in the
positions of the GCVS we are not trying to match the two remaining
semi-regulars given their smaller variation amplitude.
\section{Conclusions and outlook\label{sec.conclusions}}
We presented the catalogue of variable stars detected by the WeCAPP
project in the central parts of M31. The observations in the optical
$R$ and $I$ bands towards the bulge of M31 covering the years
2000-2003 with very good time sampling resulted in a database of over
23000 variable sources. The catalogue containing regular, semi-regular
and irregular LPVs, Cepheids, RV Tauri stars, eclipsing binary and R
Coronae Borealis candidates enlightens the rich population of variable
sources in M31. Two more papers dealing with the Fourier parameters
for the Cepheid-like stars and the WeCAPP nova catalogue are in
preparation. A fraction of the nova catalogue has already been
correlated with supersoft X-ray sources (SSSs) in M31, showing that
classical novae constitute the major class of SSSs in M31
\citep{pietsch05}.

The distribution of the identified variable sources shows an asymmetry
due to the enhanced extinction in the spiral arms projected on the M31
bulge. Assuming that the non-variable stars in the bulge behave in the
same way as the variable counterparts this asymmetry enlarges the
expected asymmetry in the microlensing signal resulting from the high
inclination of M31. Therefore theoretical calculations have to take a
modified distribution of sources (or the distribution of the
extinction present in a given band) into account.

Recently it was proposed that sequence E, first detected by
\cite{wood99} in the MACHO data towards the LMC bar, is populated by
ellipsoidal red giants \citep{soszynski04}.  \cite{wood99} found five
distinct period-luminosity sequences, of which three could be
attributed to different evolutionary phases and pulsation modes. For
two of them, the sequences D and E, it was suggested that they are
populated by contact binaries, and semi-detached binaries with an invisible
companion.  Analyzing the data of the LMC ellipsoidal candidates from
the OGLE web archive\footnote{http://ogle.astrouw.edu.pl/} and
accounting for the distance modulus of M31 proves that these sources
are below the detection limit of our survey. Therefore, we can not put
any constraints on a possible population of ellipsoidal red giants in
M31.

The more detailed examination of the group III LPVs will be subject to
future work. Until now we measured the dominant period in these stars
which are known to show multi-periodic phenomena. Future work will
include the extraction of the sub-dominant periods and the
distribution of the ratio of the different periods. Furthermore, the
distribution and possible correlations of the Fourier parameters for
these stars is an interesting and until now unstudied problem which we
can address with this dataset.

\begin{acknowledgements} 

We thank the referee Jean-Philippe Beaulieu for his comments, who
helped to improve the manuscript considerably.  The authors would like
to thank the observers and staff at Wendelstein Observatory, Otto
B\"arnbantner, Christoph Riess, Heinz Barwig, Claus G\"ossl, and
Wolfgang Mitsch, and all the staff at Calar Alto Observatory for the
extensive support during the observing runs of this project.  JF
thanks Armin Gabasch and Ulrich Hopp for stimulating discussions. This
work was supported by the {\em Sonderforschungsbereich 375-95
Astro-Particle-Physics} of the Deutsche Forschungsgemeinschaft.
\end{acknowledgements}

\appendix
\section{Survey window function and leakage function}
Effects induced by the discrete time sampling of astronomical surveys
can yield strong power in frequencies not caused by a periodic or
quasi-periodic signal. The frequencies affected are often connected to
the daily separation of observations or to observing blocks
interrupted yearly by the non-observability of the object. In case of
M31 the observation gap usually lasts from the mid of March to the
beginning of June.

A quantitative way to describe the effects of the time sampling of the
observations on the power spectrum is given by the window function
$W(\nu)$. $W(\nu)$ is defined as the Fourier transform of the sampling
function $s(t)$, where $s(t_i)$ is of constant value in case of
observations taking place at times $t_i$ and zero otherwise:

\begin{equation}
W(\nu)=\int s(t) \,e^{-2 \pi i \nu t} dt\,\,\,.
\label{eq.window_function}
\end{equation}

\begin{figure}[]
\centering
\includegraphics[width=8.3cm,angle=0]{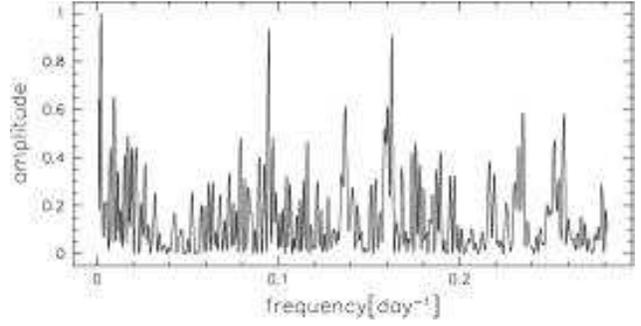}
\caption{\label{fig.window_real_f} This figure shows the window
function $W(\nu)$ present in the survey, the amplitude is normalized
to~1.  $W(\nu)$ is defined as the Fourier transform of the sampling
function s(t), $W(\nu)=\int s(t) \,e^{-2 \pi i \nu t} dt$. The main
peak of the window function has a full width half maximum
$\nu_{FWHM}=0.0013$ day$^{-1}$ which is 1.60 times larger than what
one would expect from the total duration of the observations
\citep{roberts87}.  }
\end{figure}

\begin{figure}[]
\centering
\includegraphics[width=8.3cm,angle=0]{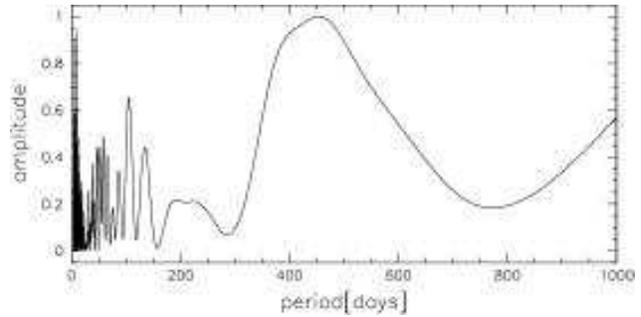}
\caption{\label{fig.window_real_p} Window function $W(\nu)$ shown in
the period representation. Clearly visible is the sampling induced
strong power in periods around 450 days which give rise to the
aliasing problems (see also Figs.  \ref{fig.window_all} and
\ref{fig.window_period}).  }
\end{figure}

In Figs. \ref{fig.window_real_f} and \ref{fig.window_real_p} we show
the window function $W(\nu)$ (in the frequency and period
representation, respectively) calculated with the Lomb algorithm for
the sampling given in Field F1. For this purpose we created a high
(order of $10^{14}$) signal-to-noise-ratio ($S/N$) light curve with
$s(t_i)= 1+\sigma(t_i)$ and a minimal noise contribution $\sigma(t_i)$
to allow for the calculation of the variance of the data (see
Eqs.~\ref{eq.pn} and \ref{eq.hsig}).

Clearly visible is the strong power in periods around 450 days (at
$\nu=2.2 \times10^{-3}$ days$^{-1}$), whereas at small periods (high
frequencies) no `aliasing' problem is expected. The full width half
maximum $\nu_{\mathrm{FWHM}}$ of the main peak of the window function
$W(\nu)$ determines the theoretical error in the period calculation

\begin{equation}
\Delta P \approx \frac{\Delta \nu}{2} P^2
\label{eq.p_error}
\end{equation}

with $\Delta \nu= \nu_{\mathrm{FWHM}}$. According to \citet{roberts87}
$\Delta \nu$ can be approximated by $\Delta \nu \approx 1/T$ with $T$
the total time span of the observations if the sampling is not to
non-uniform. The measured $\nu_{\mathrm{FWHM}} = 0.0013$ day$^{-1}$ is
1.60 times larger than this theoretical value which can be explained
by the non-uniform sampling of the observations (see
Fig.~\ref{fig.sampling}). With this value Eq.~\ref{eq.p_error} becomes
\begin{equation}
\Delta P \approx 6.5 \times \left(\frac{P}{100\, \mathrm{d}}\right)^2 \,\,\,.
\label{eq.p_error_theoretical}
\end{equation}

\begin{figure}[t!]
\centering
\includegraphics[width=0.4\textwidth,angle=0]{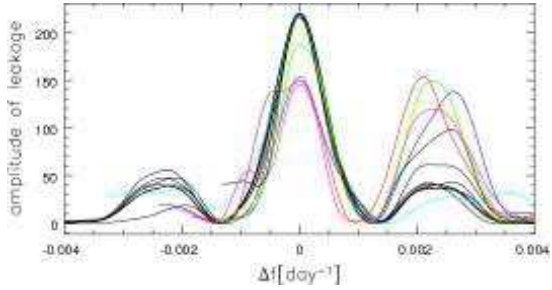}
\caption{\label{fig.window_all} Leakage function for the sampling
given in the survey for different periodic signals. The function shows
how a signal which is located at zero offset leaks into neighboring
frequencies.  We calculated this function for a signal according to
Eq. (\ref{eq.simulation}) and periods of 5, 10, 20, 50, 100, 200, 250,
300 (cyan curve), 400, 450, 500, 550, 600, and 700 days (black curves,
if not mentioned otherwise). The power in the second peak is high for
periods of 400 (blue curve), 450 (magenta curve), and 500 days (red
curve). In this period regime the power in the second peak can exceed
the power in the first peak (at zero frequency difference) resulting
in an aliasing of periods. The aliasing problem gets lower for periods
of 550 (green curve) and 600 days (yellow curve), and vanishes for
even larger periods. The aliasing problem is also evident in the
Monte-Carlo simulation of the survey.  }
\end{figure}

An alternative way to illustrate the sampling effect is given by
the `leakage function' for different periodic signals. 
The function describes how a signal which is located at zero offset `leaks' into neighboring 
frequencies. Figure \ref{fig.window_all} shows the leakage function for a periodic signal
of the form 

\begin{equation}
h_j=10^{\sin(2 \pi (t_i-t_0)/P)}
\label{eq.simulation}
\end{equation}
calculated with the Lomb algorithm for the sampling $s(t_i)$ of field
F1 and different periods $P$. The problem of high amplitudes in the
power spectrum for frequencies corresponding to periods in the range
between 350 and 550 days is apparent in this figure also.

\begin{figure}[t!]
\centering
\includegraphics[width=0.4\textwidth,angle=0]{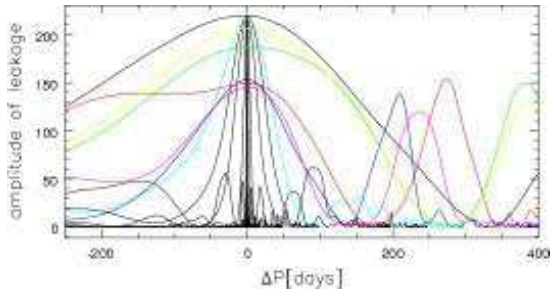}
\caption{\label{fig.window_period} Leakage function in the period
representation for the sampling given in the survey for different
periodic signals. Color coding as in Fig.\ref{fig.window_all}.  For
periods between 350 and about 550 days the power in the second peak
can exceed the power in the first peak (at zero period difference)
resulting in a possible aliasing of periods.  The resulting
displacement of the true period can be read from the x-axis.  }
\end{figure}

In Fig. \ref{fig.window_period} we show the leakage function in the
period representation.  For periods between about 350 and 550 days the
power in the second peak in the Fourier spectrum can exceed the power
in the first peak which means that the derived period is offset from
the real one by the position of the second peak.

\section{Monte Carlo simulation}
Additionally we tested the accuracy of the period determination with
Monte Carlo simulations of the survey.  We look for the dependence of
the accuracy for different mean magnitudes and amplitudes at a fixed
position, hence fixed noise contributions of the galaxy and the sky.
We choose the mean magnitude $M_R$ of the variable to be
$M_R=-1,0,1,2$ mag and shift them to the distance of M31. For each of
these mean magnitudes we change the amplitude $A_R$ of the variation
from 0.5 mag to 2 mag. The background (surface brightness of the
galaxy + sky contribution) is set to $M_{bg,R}=19$ mag per area of the
PSF. In each run we simulate $10^5$ light curves according to

\begin{equation}
F_R(t_i)=10^{-0.4(M_R+(m_R-M_R)+A_R \sin(2 \pi (t_i-t_0)/P_{in}))} \,\, .
\label{eq.sim}
\end{equation}

\begin{figure}[h!]
\centering
\includegraphics[width=0.4\textwidth,angle=0]{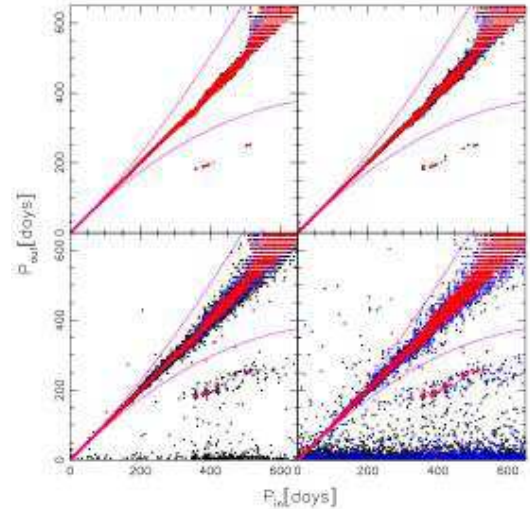}
\caption{\label{fig.inout} Result of simulations for decreasing $S/N$
of the light curves from the upper left to the lower right corner. The
mean magnitude of the source changes in steps of one magnitude from
$M_R=-1$~mag in the upper left corner to $M_R=2$~mag in the lower
right corner. Red dots show the accuracy of the period determination
for light curves with an amplitude $A_R= 2.0$~mag, blue dots
correspond to $A_R= 1.0$~mag, and black dots have $A_R= 0.5$~mag. The
apparent horizontal lines at large periods result from the finite
frequency resolution of the period finding algorithm.  Above a
threshold in $S/N$ the accuracy of the period determination seems
independent of the particular $S/N$ of the light curves. In this
regime only periods larger than 500 days have a large false detection
probability. Below the threshold the accuracy breaks down very
rapidly, independent of the period.  In magenta we show the
theoretical expected errors of the period determination according to
Eq.~\ref{eq.p_error_theoretical}.}
\end{figure}

The epochs $t_i$ are given by the sampling of field F1 (see
Fig.~\ref{fig.sampling}), periods $P_{in}$ are chosen randomly between
3 and 700 days, and $t_0$, the epoch of mean luminosity is randomly
distributed between the beginning and the end of the survey. Note,
that this also ensures that the phase is randomly chosen.  For each
simulated light curve the period is derived according to Eq.
\ref{eq.pn}. Figure \ref{fig.inout} shows the derived periods
$P_{out}$ as a function of the input period.  It is clearly seen that
the accuracy of the period determination decreases at around 500 days,
independently of the $S/N$ of the light curves.  This can partly be
explained by the sparse time sampling in the last two campaigns. On
the other hand also aliasing effects due to the yearly observing
window seem to play a role for periods between 350 days and 550 days
(see Figs. \ref{fig.window_all}, \ref{fig.window_period} and
\ref{fig.inout}). This holds for $S/N$-ratios above a certain
threshold.  Noise present in the light curves tends to level out the
different peaks in the power spectrum which results in a generally bad
performance of the period determination for light curves with low
$S/N$. For higher $S/N$ light curves the accuracy of the period
determination is good apart from the problem that periods around 400
and 500 days tend to be shifted to much smaller (i.e. around half of
the true) values.

\end{document}